\documentclass[3p,twocolumn]{elsarticle}

 \usepackage[flushleft]{threeparttable}
 \usepackage{tablefootnote}

\usepackage{graphicx}
\usepackage{dcolumn}
\usepackage{amsmath}
\usepackage{amssymb}
\usepackage{epstopdf}
\usepackage{multirow}
\usepackage{color}
\usepackage{braket}
\usepackage[alsoload=hep]{siunitx}
\usepackage{hyperref}

\usepackage{subcaption}
\usepackage{placeins}

\journal{}
\newcommand*{\MIT }{Massachusetts Institute of Technology, Cambridge, Massachusetts 02139, USA}
\newcommand*{\ODU}{Old Dominion University, Norfolk, Virginia 23529, USA}

\newcommand*{\TAU }{School of Physics and Astronomy, Tel Aviv University, Tel Aviv 69978, Israel}

\newcommand*{\NRC}{Nuclear Research Center Negev, Be'er Sheva 84190, Israel}
\newcommand*{\FSU}{Florida State University, Tallahassee, Florida 32306, USA}
\newcommand*{\UTFSM}{Universidad T\'{e}cnica Federico Santa Mar\'{i}a, Casilla 110-V Valpara\'{i}so, Chile}
\newcommand*{\GWU }{George Washington University, Washington, D.C.  20052, USA}

\begin{document}

\begin{frontmatter}

\title{The CLAS12 Backward Angle Neutron Detector (BAND) }


\author{E.P.~Segarra}
\author{F.~Hauenstein\corref{mycorrespondingauthor}\fnref{odu}}
\ead{hauenst@jlab.org}

\author{A.~Schmidt\fnref{gwu}}
\author{A.~Beck\fnref{nrcn}}
\author{S.~May-Tal Beck\fnref{nrcn}}
\author{R.~Cruz-Torres}
\author{A.~Denniston}
\author{A.~Hrnjic}
\author{T.~Kutz}
\author{A.~Nambrath}
\author{J.R.~Pybus }
\author{O.~Hen}
\address{\MIT}

\author{K.~Price\fnref{orsay}}
\author{C.~Fogler}
\author{T.~Hartlove}
\author{L.B.~Weinstein}
\address{\ODU}
\author{I.~Vega}
\author{M.~Ungerer}
\author{H.~Hakobyan}
\author{W.~Brooks}
\address{\UTFSM}
\author{E.~Piasetzky}
\author{E.~Cohen}
\author{M.~Duer}
\address{\TAU}
\author{I.~Korover}
\address{\NRC}
\author{J.~Barlow}
\author{E.~Barriga}
\author{P.~Eugenio}
\author{A.~Ostrovidov}
\address{\FSU}

\cortext[mycorrespondingauthor]{Corresponding Author}

\fntext[odu]{Also at: \ODU}
\fntext[gwu]{Current address: \GWU}
\fntext[nrcn]{Also at: \NRC}
\fntext[orsay]{Current address: Institut de Physique Nucl\'{e}aire, IN2P3-CNRS, F-91406 Orsay, France}
\date{\today}
\begin{abstract}
 The Backward Angle Neutron Detector (BAND) of CLAS12 detects neutrons
  emitted at backward angles of $155$\si{\degree} to $175$\si{\degree},
  with momenta between $200$ and $600$ \si{\MeV/\clight}. It is
  positioned 3-\si{\meter} upstream of the target, consists of $18$
  rows and $5$ layers of $7.2$-\si{\centi\meter} by $7.2$-\si{\centi\meter} 
  scintillator bars, and read out on both ends by PMTs to
  measure time and energy deposition in the scintillator
  layers. Between the target and BAND there is a 2-\si{\centi\meter}
  thick lead wall followed by a 2-\si{\centi\meter} veto layer to
  suppress gammas and reject charged particles. 
  
This paper discusses the component-selection tests and the
  detector assembly. Timing calibrations (including offsets and time-walk) were performed using a novel pulsed-laser calibration system, resulting in
  time resolutions better than $250$~\si{\pico\s} (150~\si{\pico\s}) for energy depositions above 2 MeVee (5 MeVee). Cosmic rays and a variety of radioactive sources were used to 
  calibration the energy response of the detector. Scintillator bar
  attenuation lengths were measured. The time resolution results in a neutron momentum
  reconstruction resolution, $\delta p/p < 1.5$\% for neutron momentum
  $200\le p\le 600$ MeV/c. Final performance of the BAND with CLAS12
  is shown, including electron-neutral particle timing spectra and a
  discussion of the off-time neutral contamination as a function of
  energy deposition threshold.
  
\end{abstract}

\begin{keyword}
CLAS12; time of flight; plastic scintillator; ultrafast neutrons
\end{keyword}
\end{frontmatter}


\setcounter{footnote}{0}
\renewcommand{\thefootnote}{\alph{footnote}}
\section{Introduction}

CLAS12 (12-GeV CEBAF Large Acceptance
Spectrometer)~\cite{Burkert:2020akg} in Hall B of the Thomas Jefferson
National Accelerator Facility (JLab) is a multi-purpose spectrometer
used to detect charged and neutral particles emitted in high-energy
electron scattering reactions.  Tagged deep inelastic scattering studies
 require the detection of backward-angle recoil
neutrons with momenta above \SI{200}{\mega\eVperc}~\cite{band-proposal}.
This is achieved by a combination of the Central Neutron Detector
(CND)~\cite{Chatagnon:2020lwt} and the Backward Angle Neutron Detector
(BAND). The CND covers angles from $40$\si{\degree} to
$120$\si{\degree} while BAND covers from $155$\si{\degree} to
$175$\si{\degree}.

\begin{figure}[t!]
	\centering
	\includegraphics[width=0.49\textwidth]{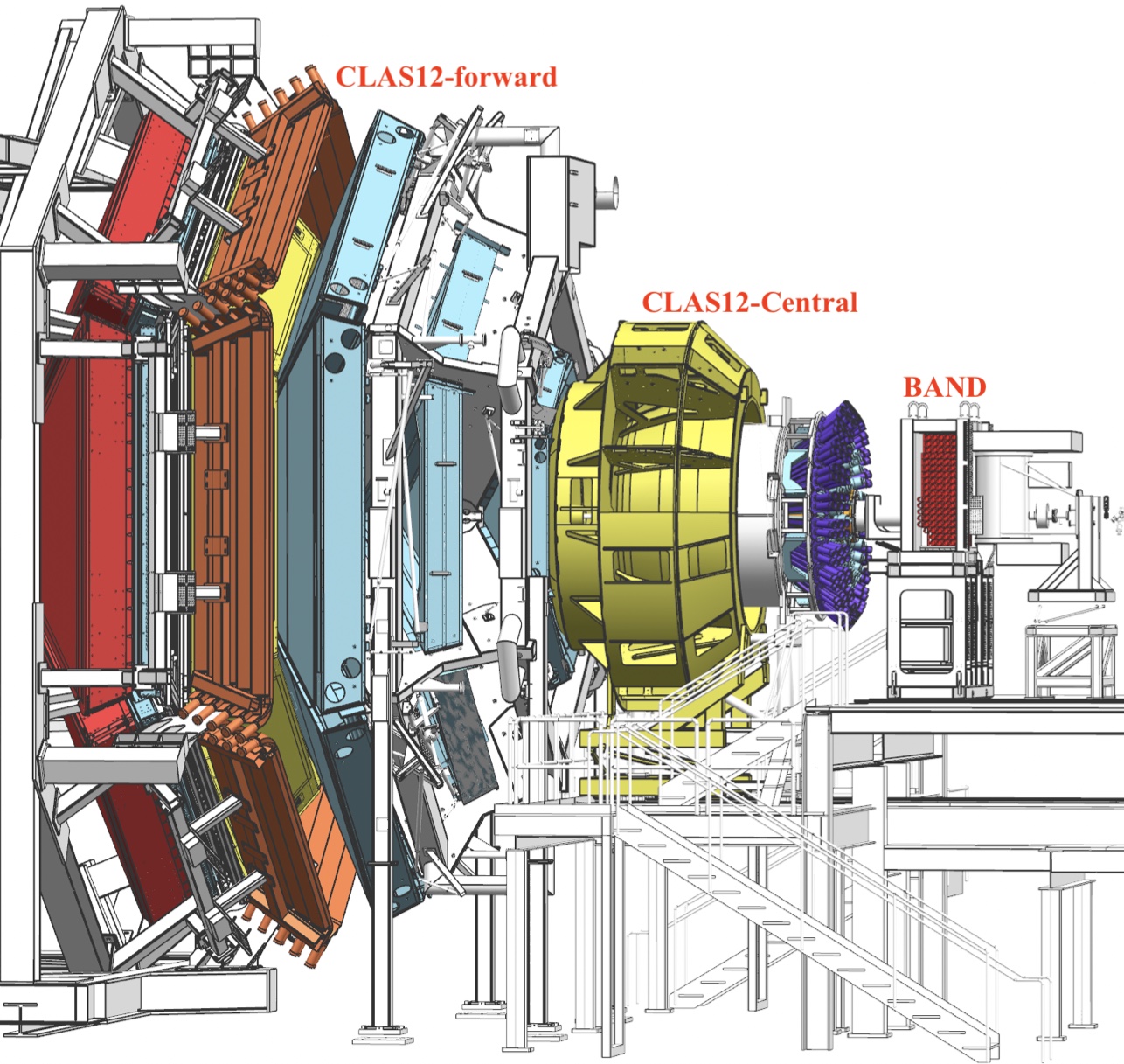}
        \caption{CLAS12 and BAND in Hall B at Jefferson Lab. The
          electron beam is incident from the right side. BAND is
          marked in red. The overall detector system is roughly 20
          \si{\meter} in scale along the beam axis. }
		\label{fig:clas12band}
\end{figure}

BAND was designed to measure neutrons with
momenta of $200 - 600$ \si{\MeV/\clight}, with an average detection efficiency
of $35$\%, and with a momentum reconstruction resolution better than
$1.5\%$. The detector is based on scintillator bars with PMT readout on both bar ends. BAND was
installed in January 2019 and has since taken data in coincidence with
CLAS12. Fig.~\ref{fig:clas12band} shows a drawing of BAND located
upstream of CLAS12 in Hall B.

This paper describes the design, operation and calibration of the BAND.
Section 2 presents the required time resolution, efficiency and
geometry, the resulting design and layout, the selection of detector
components, and results from comparative measurements of different
PMTs and magnetic shields. Section 3 presents the performance of the
detector after its installation in Hall B. It describes cosmic-ray and
laser calibrations ~\cite{band-laser} as well as the measured
time-of-flight (ToF) resolutions and neutron-photon separation from
$10.6$-\si{\GeV} electron-deuteron data. Section 4 summarizes the
results.


\section{Design of the backward angle neutron detector}
The major considerations for the BAND design were the constraints on
geometry and areal coverage in Hall B, the time-of-flight resolution
required, and neutral particle identification.  To achieve the physics
of interest in BAND~\cite{band-proposal}, neutron ToF
resolutions below $300$ \si{\pico\second} at an energy deposition
threshold of $\sim 2$ MeVee (MeV-electron-equivalent) are required. 
This threshold is later optimized using neutron signal-counts-above-background,
measured ToF resolutions and neutron efficiencies as a function of minimum energy deposition.
Neutral particles are
identified by using a thin $2$-\si{\centi\meter} veto layer for
charged-particle identification between the target and the first
active layer of BAND. Photons are suppressed by a 2-\si{\centi\meter}
thick lead wall placed downstream of the veto layer. Neutron-photon
discrimination is achieved via ToF relative to the electron scattering
time measured by CLAS12. Out-of-time random neutron and photon
contamination in a signal region can be reduced, relative to the signal strength, 
by optimizing the minimum energy deposition of particles in the bars (see discussion in {\it{Neutron identification}}).

The following subsections describe the BAND geometry, the individual components,
including the bench measurements that informed their selection, the photon
shielding, the laser calibration system, the electronics and data acquisition, 
and the final assembly of the detector.

\subsection{Geometry}
\begin{figure}[tb]
	\centering
		\includegraphics[width=0.40\textwidth]{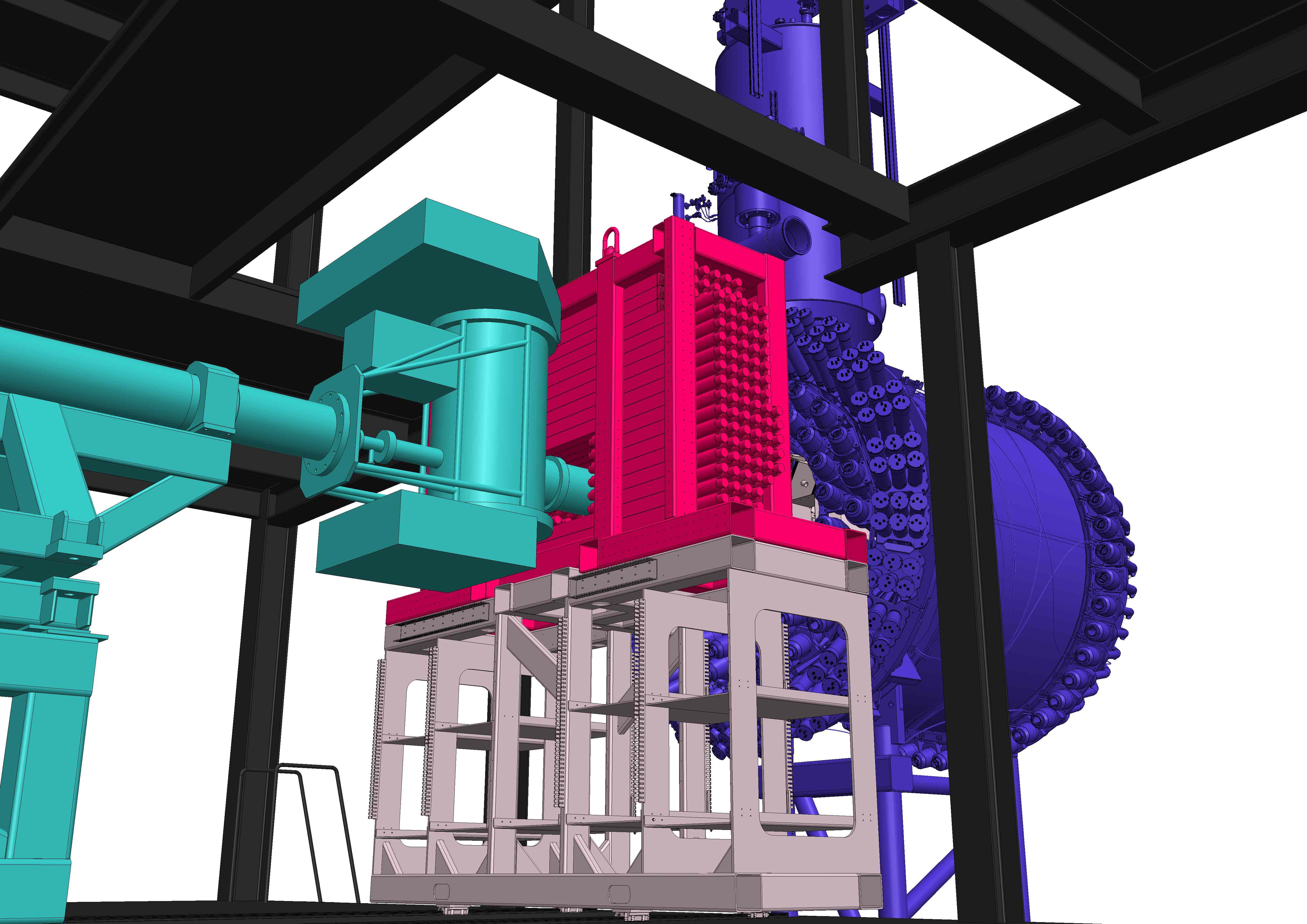}
		\vspace{0.5cm}
	    \caption{BAND and its close surroundings: the
                  cryotarget and beam line components (cyan),  BAND
                  (red), the central detector region of CLAS12
                  (blue), the space frame (black), and the central detector support cart (grey). The beam is coming from the left side. The
                  target is located at the center of the CLAS12 central detector.}
		\label{fig:bandtarget}
		
\end{figure}

The design of the BAND geometry balanced the goal of maximizing acceptance for backward angle neutrons
with the space contraints in hall (see Fig.~\ref{fig:bandtarget}). BAND was designed to cover the region
between the Central Detector and the target support, where the flight path of the neutrons had minimal
material obstruction. In order to be installed on the support cart for the central tracking detectors, BAND
had to be lowered down through an existing opening in the CLAS12 support scaffolding (referred to as the space
frame, shown in black in Fig.~\ref{fig:bandtarget}). This required that BAND fit within a box 3-\si{\meter} wide, 1.5-\si{\meter} high, and 1-\si{\meter} deep, and include a hole in the middle for the beam pipe.

\begin{figure}[tb]
	\centering
			\includegraphics[width=0.48\textwidth]{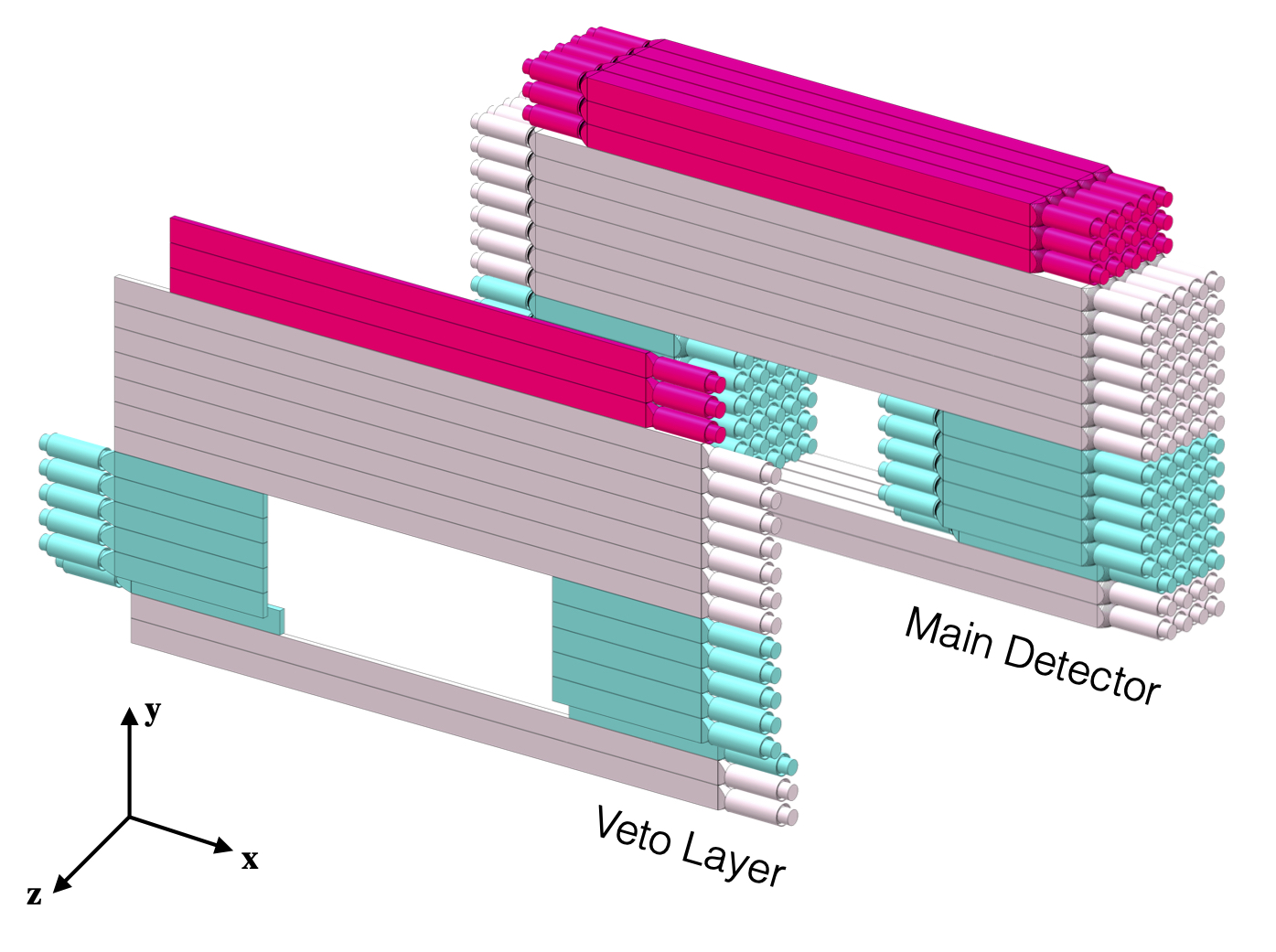}
            \caption{BAND design of bars in the main
                   detector and the veto layer. The 164-\si{\centi\meter} long bars are shown in
                          red, the 202-\si{\centi\meter} bars in white
                          and the short, 51-\si{\centi\meter} bars in cyan. The position of
                          the PMTs for each bar is also shown. The beam direction corresponds to the $z$-axis.   }
		\label{fig:design}
\end{figure}

To maximize the active detector volume within these constraints, we
chose to use rectangular scintillator bars stacked (in $x$-$y$), perpendicular to
the beam direction ($z$), see Fig.~\ref{fig:design}.  The cross
section of each scintillator bar determines our position granularity
in $y,z$ (vertical and longitudinal directions relative to the beam
pipe), and should be comparable to our time-resolution-dependent position
resolution along the bar (in $x$). Given our time-resolution specification of
$300$ \si{\pico\second} (discussed below), $7.2$
\si{\centi\meter} by $7.2$ \si{\centi\meter} scintillator bars were
chosen to optimize fiducial volume, granularity, and cost.

The active detector area (excluding the veto layer) consists of 116 scintillator bars, arranged
in 5 $z$-layers with 18 vertically stacked rows of bars. The arrangement of the
bars is shown in Fig.~\ref{fig:design}. The bottom three rows each have only
four layers due to obstructions from the cart below BAND (see grey cart in Fig.~\ref{fig:bandtarget}).

Three different bar lengths are used: 15 bars of length 164 \si{\centi\meter}, 43 bars of length 202 \si{\centi\meter}, and 58 bars of length 
51 \si{\centi\meter}. The shorter bars were used in the vicinity of
the beam pipe. All bars have light guides attached to both sides, and the bars
are read out by 51 \si{\milli\meter} PMTs. The PMTs are either Hamamatsu R7724~\cite{pmtR7724} or Electron Tubes (ET) 9214KB~\cite{pmt9214} (see Table~\ref{tab:geometry}).

The veto layer, which is installed on the downstream face of BAND
(i.e., between BAND and the target),
consists of 24 scintillator bars with a cross section of $2$
\si{\centi\meter} by $7.2$ \si{\centi\meter}. The veto bars have the
same lengths as the corresponding BAND bars, such that the whole
downstream surface is covered by the veto, see
Fig.~\ref{fig:design}. Each veto bar is only read out on one side by a
single ET 9954KB~\cite{pmt9954} 51 \si{\milli\meter} PMT.

Each bar has an assigned sector and layer number. The layer number
corresponds to the position along the beam line (in $z$) starting
from upstream (1) to downstream (5) for the active area and (6) for
the veto bars. The sector numbering follows from top to bottom. Sector 1 
consists of three rows of 164-\si{\centi\meter} long bars (red bars in Fig.~\ref{fig:design}). 
Sector 2 has seven rows of 202-\si{\centi\meter} long bars, below Sector 1. Sector 5 also 
consists of the 202-\si{\centi\meter} long bars, two rows on the bottom of the detector. 
The white bars in Fig.~\ref{fig:design} show the locations of Sectors 2 and 
5 in each layer. The short 51-\si{\centi\meter} bars to the left of the beam hole (negative $x$) are in Sector 3, while the ones on the right side (positive $x$) 
are in Sector 4 (both are shown in cyan in Fig.~\ref{fig:design}). 
Table~\ref{tab:geometry} summarizes the geometry parameters 
and PMTs for each sector and layer.

\begin{table}[t]
\caption{Parameters for bars and PMTs for the different BAND sectors and layers.}
\centering
\begin{tabular} {l  l  m{4em}} \hline
 &  Dimensions ($w\times h$) & PMT \\ \hline\hline
\multicolumn{3}{l} {Sector 1,  $L = 164$ \si{\centi\meter}} \\ \hline
Layer $1 - 5$  & 7.2$\times$7.2 \si{\centi\meter\squared} & R7724  \\
Layer 6  & 2$\times$7.2 \si{\centi\meter\squared} & 9954KB  \\
\hline
\multicolumn{3}{l} {Sector 2, $L = 202$ \si{\centi\meter}} \\ \hline
Layer $1 - 5$  & 7.2$\times$7.2 \si{\centi\meter\squared} & R7724  \\
Layer 6  & 2$\times$7.2 \si{\centi\meter\squared} & 9954KB  \\
\hline
\multicolumn{3}{l} {Sector 3 / 4, $L = 51$ \si{\centi\meter}} \\ \hline
Layer $1 - 4$  & \multirow{2}{*}{7.2$\times$7.2 \si{\centi\meter\squared}} & 9214KB \\
Layer 5 & & R7724 \\
Layer 6  & 2$\times$7.2 \si{\centi\meter\squared} & 9954KB  \\
\hline
\multicolumn{3}{l} {Sector 5, $L = 202$ \si{\centi\meter} } \\ \hline
Layer $1 - 4$\tablefootnote{Layer 5 is missing due to obstructions from the central detector support cart below BAND}  & 7.2$\times$7.2 \si{\centi\meter\squared} & R7724  \\
Layer 6  & 2$\times$7.2 \si{\centi\meter\squared} & 9954KB  \\
\hline
\end{tabular}
\label{tab:geometry}
\end{table}

\subsection{Components}
The timing resolution of the scintillator bars is affected by many
factors, and each component was optimized considering both cost and
 design constraints.
We selected Bicron BC-408~\cite{scint-mat-ref} scintillant for its
light output, time response and attenuation length. 
The bulk attenuation length of 380~\si{\centi\meter} is much longer
than the length of the BAND bars. To enhance reflectivity, we wrapped
each bar with 3M Enhanced Specular Reflector foils~\cite{3MESR}. 
In order to select the optimal PMTs and thickness of magnetic-shields, we 
bench-tested a variety of options described in the following text.

\begin{figure}[tb]
	\centering
		\includegraphics[width=0.48\textwidth]{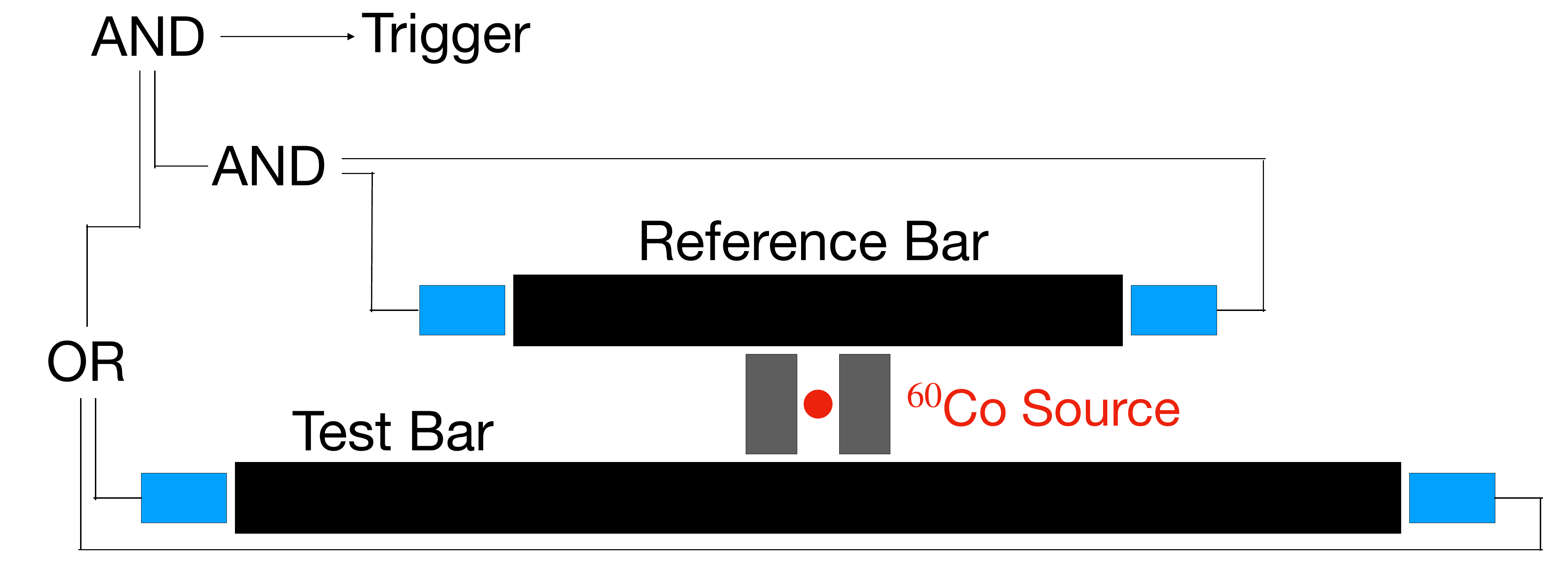} \\
		\includegraphics[width=0.48\textwidth]{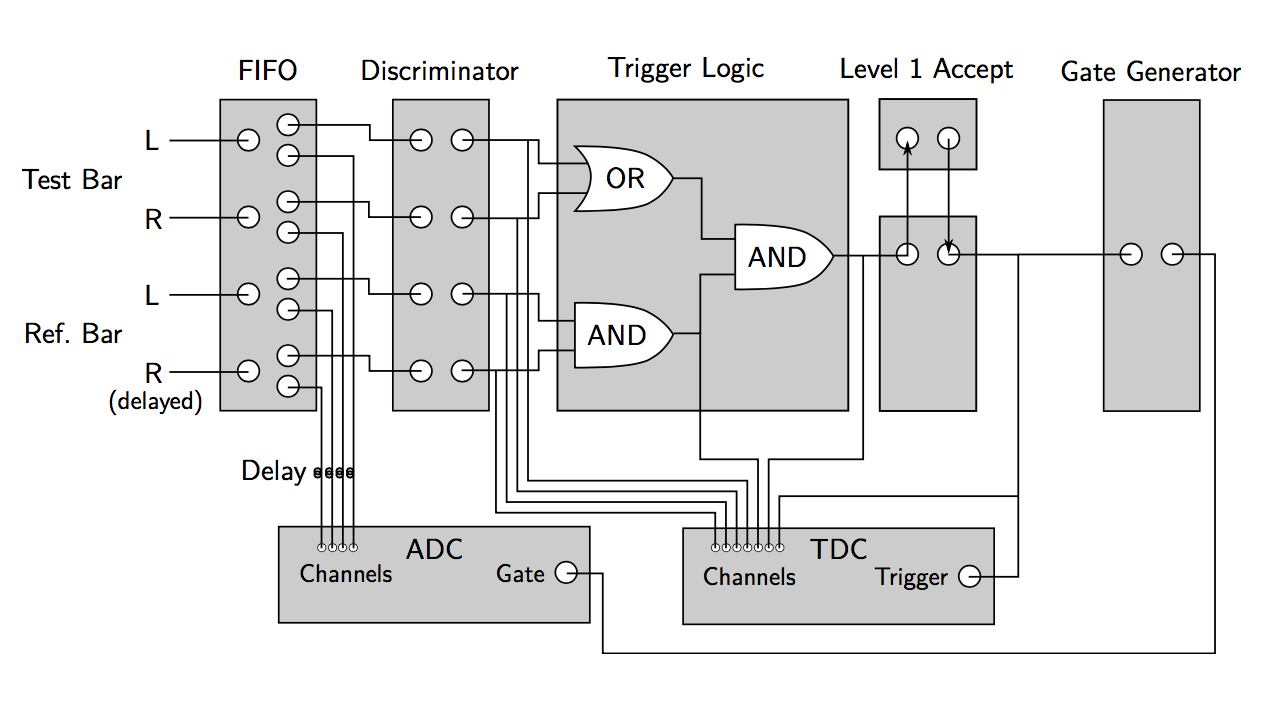}
	\caption{Schematic of the bench test setup (top) and its
          read-out system (bottom) to measure PMT time resolutions. }
	\label{fig:test_stand_setup}
\end{figure}

\subsubsection{Bench Measurements}

Bench measurements were used to guide the PMT and magnetic shielding
selection.  A diagram of our test setup and electronics is shown in
Fig.~\ref{fig:test_stand_setup}, featuring a coincidence setup between
a ``test'' scintillator bar and a ``reference'' scintillator bar.
Each bar has two PMTs coupled to its ends; the test bar has the PMTs
whose time resolution we wish to study.  Both bars are wrapped in an
optical reflector, and placed in a dark-box.  The reference bar was
kept fixed during all measurements. Five different
51-\si{\milli\meter} PMTs were tested (see
Table~\ref{tab:tests},~\cite{hamapmts}, and~\cite{pmt9214}), assembled
on 200- to 250-\si{\centi\meter} long bars.

The signal for our measurement is given by a $^{60}$Co source which
is placed between the two bars.  The $^{60}$Co source yields two
$\gamma$ rays with energies of 1.17 and 1.33 \si{\mega\electronvolt}.
The two $\gamma$-rays were collimated by two lead bricks to ensure
they each hit a specific location along each bar, allowing us to study individual
PMT time resolution and measured energy as a function of hit location.

\begin{table}[t]
	\caption{Bench test configurations. Length refers to the length of the scintillator bar. $\sigma$ is the time
          resolution for 2-\si{\mega\electronvolt} central-equivalent energy deposit in the middle of the bar
          extrapolated from data. The 9214KB PMT~\cite{pmt9214} is manufactured by ET-Enterprises and the other PMTs are manufactured by Hamamatsu~\cite{hamapmts}.}
    \centering
	\begin{tabular}{ m{5em}   m{3em}   m{3em} }
		\hline
			PMT & Length (cm) & $\sigma$ (\si{\pico\second})\\
		\hline\hline
			R7724 &  200 &  $\sim240	$	\\
			R7724-100 & 250 & $\sim210$ 		\\
			R13089 & 250 & $\sim210	$		\\			
			R13435 & 200 & $\sim310$			\\		
			9214KB & 200 & $\sim260$			\\
		\hline
	\end{tabular}
	\label{tab:tests}
\end{table}
\begin{figure}[tb]
	\centering
		\includegraphics[width=0.48\textwidth]{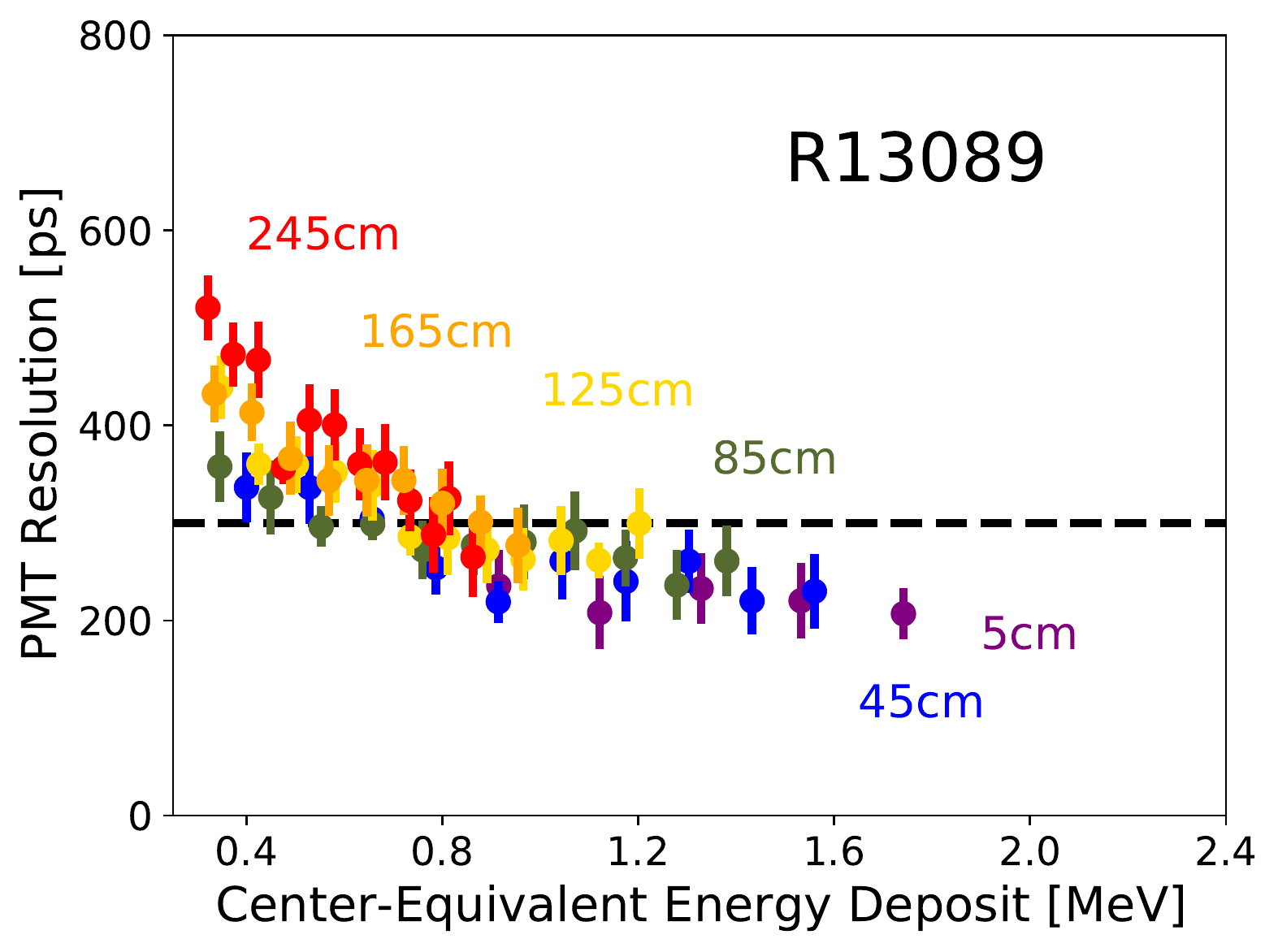}
		\caption{PMT time resolution as a function of
                  center-equivalent energy deposit for different
                  placements of the $^{60}$Co source along the bar for
                  Hamamatsu R13089 PMT. The labels indicate the
                  distance of the source to the PMT. }
	\label{fig:test_stand_posdep}
\end{figure}

Measurements with different $^{60}$Co
source locations were combined by using the measured attenuation in the bar to convert
the energy measured by the PMT to the ``center-equivalent energy
deposit,'' the energy that would have been deposited at the center of
the bar to give the same measured energy.  For example, by placing the
source close to one PMT one can achieve a center-equivalent energy
deposit significantly greater than the 1 \si{\mega\electronvolt} $^{60}$Co Compton edge.

Fig.~\ref{fig:test_stand_posdep} shows the PMT time resolution as a
function of center-equivalent energy deposited for different source placements. Each individual data set explores the response of the PMT over the entire Compton distribution for a single source placement. Since the individual data sets are consistent within error bars,
the measurements were combined to cover a center-equivalent energy deposition range
up to 1.8 \si{\mega\electronvolt}.

\begin{figure}[tb]
	\centering
		\includegraphics[width=0.48\textwidth]{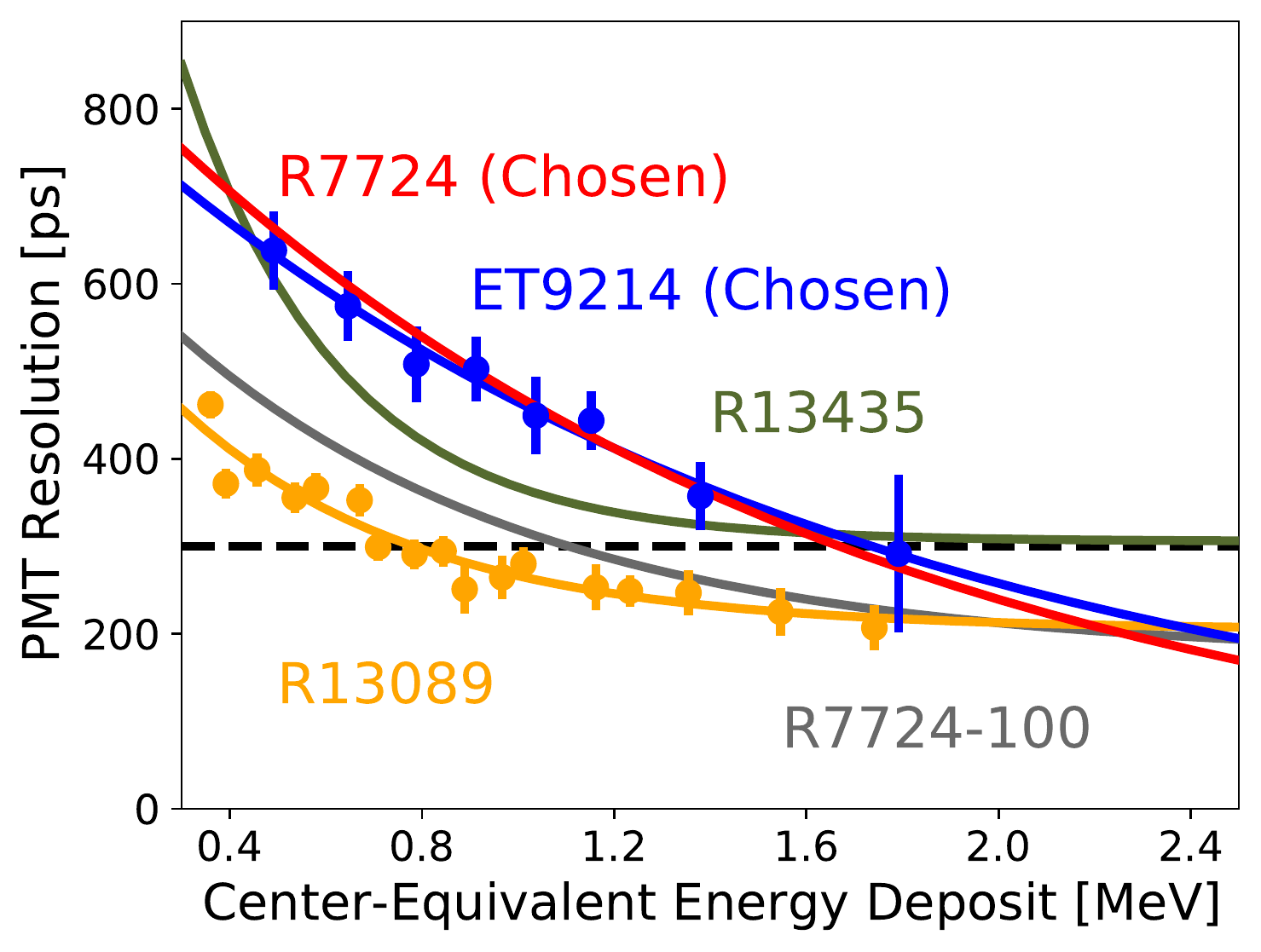}
		\caption{Single-PMT time resolution as a function of center-equivalent energy deposit for the five PMTs
                  tested. For each PMT, data, such as that of Fig.~\ref{fig:test_stand_posdep}, are combined to a single set and weighted averages are taken to reduce error given the density of points. 
                  The lines are fits to data (only the blue  and yellow data points are shown). The functional form used, $\sigma_{PMT}=a e^{-b \cdot E_{dep}} + c$, 
                  reflects that at sufficiently high energy deposition, PMT resolution becomes constant. At 2
                  \si{\mega\electronvolt} energy deposition all PMTs except R13435 achieve a time resolution below 300 \si{\pico\s}.}
          	\label{fig:test_stand_results}
\end{figure}

Fig.~\ref{fig:test_stand_results} shows the measured single-PMT time
resolution for the different test configurations.  The lines indicate
fits to data extrapolated beyond 1.8 \si{\mega\electronvolt}.  All
PMTs except the R13435 met the timing-resolution design specification
of 300 \si{\pico\s} at 2-\si{\mega\electronvolt} energy deposition
(see Table~\ref{tab:tests}).  To reduce costs, a combination of
Hamamatsu R7724~\cite{pmtR7724} and Electron Tube
9214KB~\cite{pmt9214} PMTs for BAND were selected. For the veto layer,
available Electron Tube 9954KB PMTs were used. See
Table~\ref{tab:pmts} for details of the selected PMTs.

\begin{figure}[tb]
	\centering
		\includegraphics[width=0.48\textwidth]{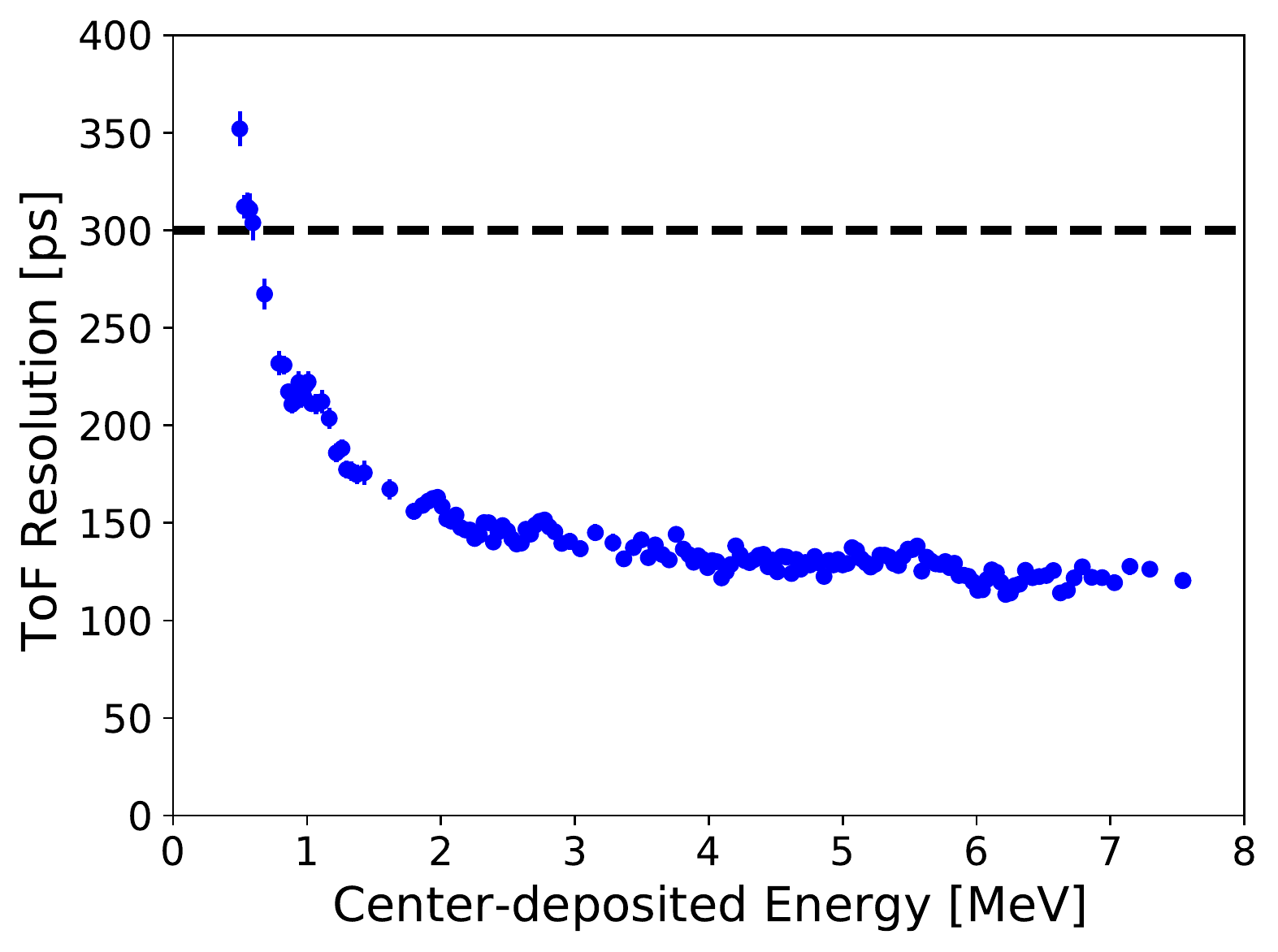}
		\caption{Time resolution as a function of center-deposited energy for one bar with R7724 PMTs using the
                  BAND laser system after installation. The energy deposition is the geometric mean of both PMTs at the
                  end of the bar, and is corrected for attenuation effects. The deposition is in the center of the bar due to the installation of the fiber there.}
         \label{fig:resolution-laser}
\end{figure}

After BAND was assembled and installed, the time resolution of each bar was measured using the laser calibration system (see Sec.~\ref{sec:laserystem} and~\cite{band-laser}). The time resolution as a function of the energy deposit in the center of the bar
is shown in Fig.~\ref{fig:resolution-laser} for a representative 202-\si{\centi\meter} bar 
with R7724 PMTs. The resolution is below 150
\si{\pico\s} from 2 to 8 \si{\mega\electronvolt} center-deposited energy, 
indicating the good performance of the bars and PMTs. One should note that 
Fig.~\ref{fig:resolution-laser} is the bar ToF resolution, while~\ref{fig:test_stand_results} is the PMT resolution. In Fig.~\ref{fig:resolution-laser}, the energy deposition is measured from both PMTs (reconstructed as the 
geometric mean and corrected for attenuation).

\begin{table*}[tbh]
\caption{Properties of the PMTs used in BAND. }
\centering
\begin{tabular} { l  l  l l } \hline
 &  R7724~\cite{pmtR7724} & 9214KB~\cite{pmt9214}& 9954KB~\cite{pmt9954} \\ \hline\hline
Company                                  & Hamamatsu       & ET           & ET \\
Diameter                                           & 51 \si{\milli\meter}                         & 51 \si{\milli\meter}                        & 51 \si{\milli\meter} \\ 
Dynode stages                                  & 10                           & 12                              & 12 \\
Spectral response                            & 300 $-$ 650 \si{\nano\meter} & 290 $-$ 630 \si{\nano\meter}  & 290 $-$ 680 \si{\nano\meter}  \\ 
Max. wavelength emission               & 420 \si{\nano\meter}  & 350 \si{\nano\meter}  & 380 \si{\nano\meter}  \\ 
Quantum eff. maximum   & 26\% & 30\% & 28\% \\ 
Gain                                                    & 3.3$\times$10$^6$ & 3.0$\times$10$^7$ & 1.8$\times$10$^7$ \\ 
Max. anode current rating               & 200 \si{\micro\ampere} & 100 \si{\micro\ampere} & 100 \si{\micro\ampere} \\ 
Anode dark current                         & 6 \si{\nano\ampere} & 4 \si{\nano\ampere} & 8 \si{\nano\ampere} \\ 
Anode pulse rise time                      & 2.1 \si{\nano\second} & 2.0 \si{\nano\second} & 2.0 \si{\nano\second} \\ 
Electron transit time                        & 29 \si{\nano\second} & 45 \si{\nano\second} & 41 \si{\nano\second} \\
\hline
\end{tabular}
\label{tab:pmts}
\end{table*}

\subsubsection{Light Guides \label{LightGuide}}
\begin{figure}[tb]
	\centering
	\includegraphics[width=0.23\textwidth]{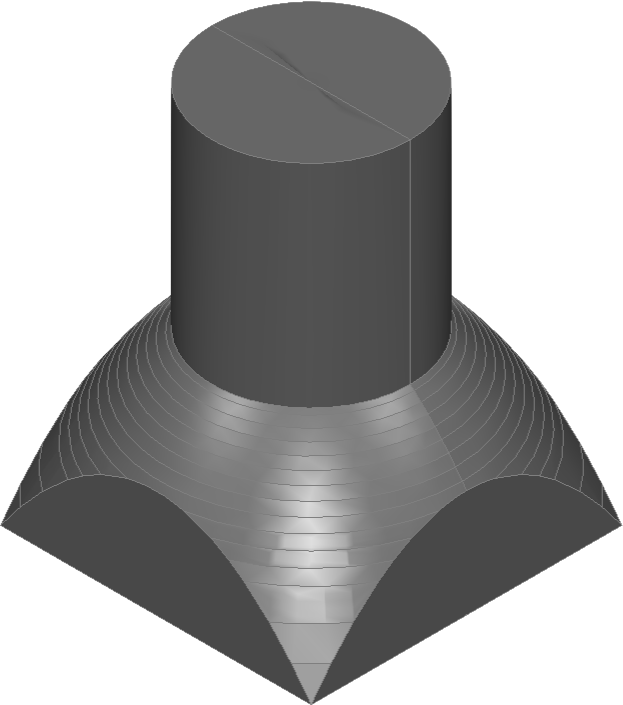}
		\includegraphics[width=0.18\textwidth]{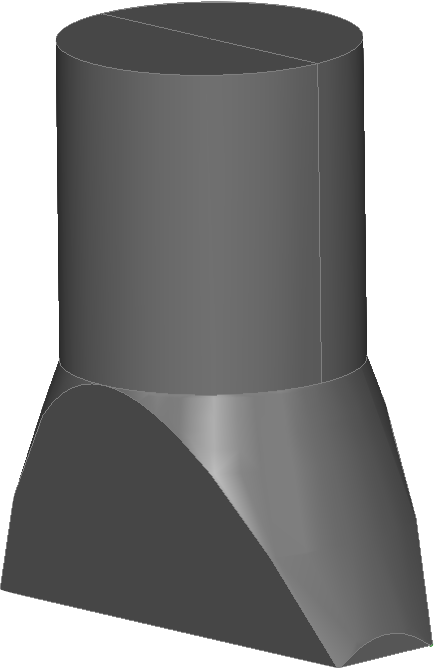}
	\caption{CAD drawing of the light guides for the main detector (left) and the veto bars (right). The top parts are connected to the PMTs while the lower parts are connected to the bars.}
	\label{fig:lightguides}
\end{figure}

The light guide design was constrained by the available space, the
sizes of the scintillator bars and PMTs, and the length of the
mu-metal shielding. The design utilizes a modified Winston cone to
optimize light collection by concentrating light from a large area at
the scintillator onto a smaller active area of the photomultiplier.
CAD drawings of the light guides for the main detector and the veto
bars are shown in Fig.~\ref{fig:lightguides}. In total, they are 8.9
\si{\centi\meter} long. Each light guide has a 4.9-\si{\centi\meter}
long cylindrical section with a 4.6-\si{\centi\meter} diameter, which
is connected to the active photocathode area of the 51-\si{\milli\meter} diameter PMTs.  The cylindrical length allows for
the magnetic field shielding to extend more than 51 \si{\milli\meter}
(i.e., more than a PMT diameter) beyond the photocathode (see next
section). The other end of the light guide matches the cross section
of the scintillator. The light guide is glued to the scintillator bar
with a DYMAX UV curing glue~\cite{uvglue} and attached to the PMT with
MOMENTIVE RTV615 silicone rubber compound~\cite{softglue}. The light
guides were designed, optimized, and manufactured by Florida State
University.

\subsubsection{PMT Magnetic shielding}

The fringe field of the CLAS12 solenoid magnet
\cite{Fair:2020yfx} is between 20 and 120 \si{\gauss} at the location of
BAND, requiring all PMTs to have magnetic shielding. The transverse
field (perpendicular to the axis of the PMT) in the PMT region ranged from 10 to
  110 \si{\gauss} and the longitudinal field ranged from 10 to 50 \si{\gauss}.

  A dedicated test stand was used to examine different thicknesses of
  passive magnetic shielding. Two 56-turn Helmholtz coils, with a
  radius of 0.5 \si{\meter}, separated by 1 \si{\meter}, were
  constructed. A prototype detector was also used, consisting of a
  2-\si{\centi\meter} thick, 5-\si{\centi\meter} diameter scintillator
  disc, coupled to two 7-\si{\centi\meter} long, 5-\si{\centi\meter}
  diameter cylindrical acrylic light guides and two PMTs. The detector
  was light-proofed with Tedlar\textregistered{} and stabilized on a
  simple plastic stand at the center of the Helmholtz coils.

  As a baseline, responses to $^{90}$Sr and $^{137}$Cs sources were measured with zero magnetic field and no shielding. The sources were placed either 0 or 5 \si{\centi\meter} away from the scintillant. PMT event rates, time differences between the left and right PMT signals, and peak energies were measured at nominal operating voltages.
\begin{figure}[tb]
	\centering
			\includegraphics[width=0.47\textwidth]{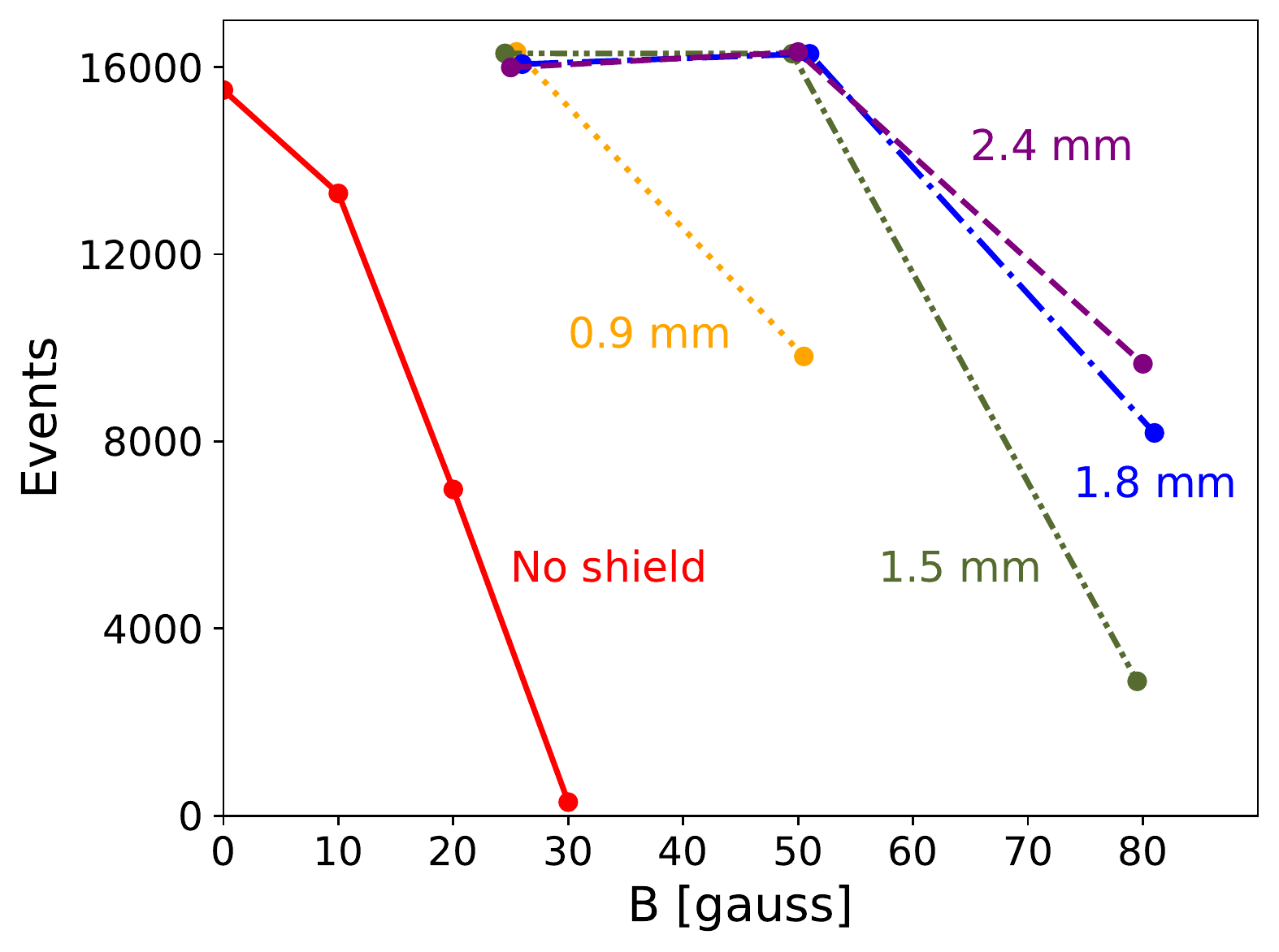}
	\caption{Magnet shielding test results with a $^{90}$Sr source, a longitudinal field. The shields were placed to cover 5 \si{\centi\meter} beyond the PMT photocathode. Event rates with increasing field and no shielding are shown in red. Results with 0.9, 1.5, 1.8, and 2.4 \si{\milli\meter} mu-metal thickness are shown in yellow, green, blue, and purple, respectively.}
	\label{fig:shielding_results}
\end{figure}  
  
  The same quantities were subsequently measured with different
  shielding configurations at different longitudinal and transverse
  magnetic fields, up to 80 \si{\gauss} (limited by heat dissipation in the
  Helmholtz coils).

 Mu-metal shielding of thicknesses of 0.9, 1.5, 1.8,
  2.4 and 3 \si{\milli\meter} were tested, as well as two soft-iron pipes. The magnetic field, amount of detector covered by the
  shielding, the thickness of the shielding, and the angle of the detector to the field were
  all varied.
  
In Fig.~\ref{fig:shielding_results}, the event rate from a $^{90}$Sr source as a function of magnetic field strength is shown for different thickness of mu-metal shields and a reference measurement without any shield. The shield was positioned to cover 5 \si{\centi\meter} beyond the photocathode. 
The longitudinal magnetic field degraded the PMT performance
the most.  While the 1.5-\si{\milli\meter} mu-metal shield was not
sufficient, the 1.8-\si{\milli\meter} mu-metal shield maintained PMT
performance up to 80 \si{\gauss}.  The soft-iron pipes performed worse than the
1.5-\si{\milli\meter} mu-metal (not shown in Fig.~\ref{fig:shielding_results}). The magnetic shields performed best when they
extended 5 \si{\centi\meter} (one diameter) beyond the PMT photocathode.  In order to
be conservative, 2.6-\si{\milli\meter} thick mu-metal shields were selected.

\subsubsection{PMT and bar quality test measurements}
Before the final assembly of BAND (see Section~\ref{sec:assembly}), all PMTs and bars have been tested for quality assurance. These measurements were conducted with a picosecond pulsed UV-LED.

The test setup for each PMT had a 6.4-\si{\centi\meter} long scintillator disc attached between the LED and the PMT inside of a light-tight dark box. Gains and time resolutions for each PMT were measured. The PMTs were then matched in pairs to be then glued to a bar. Pairs were determined by minimizing gain different and time resolution difference between the PMTs of a pair.

For each bar the attenuation length was measured by moving the LED pulser along the bar. This was done without a wrapping material around the bar and with no light guides attached, in a light tight black-box. The bars with the worst attenuation length were selected as back-up bars.

\subsection{Lead shielding}
To minimize background from backward-going photons, a
lead wall was installed on the downstream face of BAND (between the CLAS12 target
and BAND). This lead wall consists of individual 2-\si{\centi\meter}
thick blocks stacked in front of the veto layer. Each block is covered
on both sides with an aluminum layer to safely handle. The
lead wall can be seen in the final assembly photographs shown in
Fig.~\ref{fig:band_downstream}.

\subsection{Laser Calibration system}
\label{sec:laserystem}
\begin{figure}[tb]
	\centering
		\includegraphics[width=0.43\textwidth]{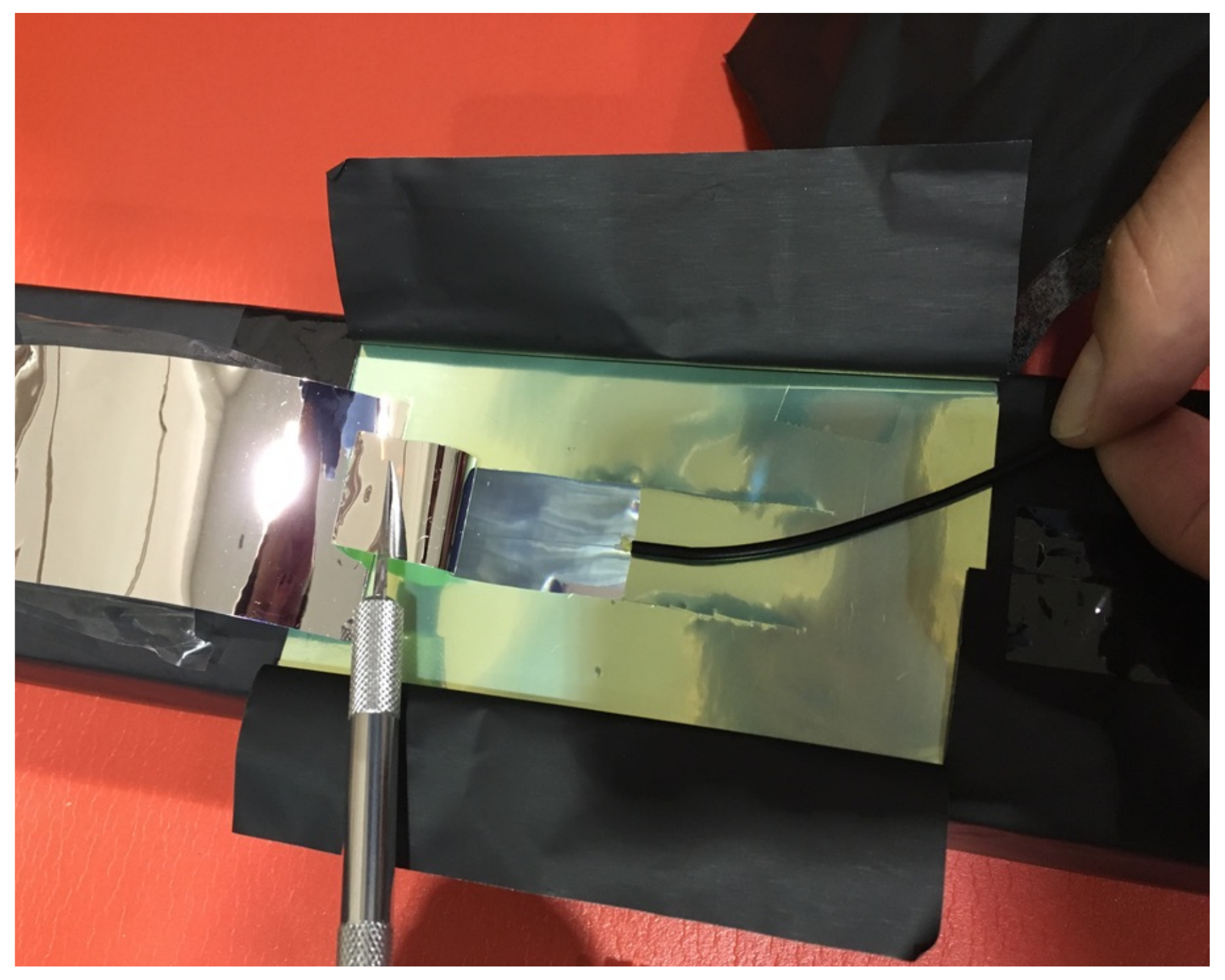}
	\caption{Connection of a fiber to bar after UV-cure gluing. The fiber is positioned in the center of the bar and is covered by both layers of wrapping material, ESR~\cite{3MESR} and Thorlabs Aluminium foil~\cite{thorlabsfoil}.}
	\label{fig:pic-fiberonbar}
\end{figure} 
A UV laser system to calibrate BAND and monitor its
performance was implemented during data-taking. 
The design and performance of the system is fully 
described in Ref.~\cite{band-laser}. 

The main component of the system is a picosecond pulsed diode laser 
(Teem Photonics STV-01E-140,~\cite{teem_laser}) with a wavelength 
of 355 \si{\nano\meter}. The laser can be triggered externally between 
$10-4000$ \si{\hertz}. The light is only transported within fibers. Further components of the system are a mode scrambler, a 90:10 splitter whose outputs are connected to a reference photodiode (10\% output) and a variable optic attenuator~\cite{attenuator} (90\% output). The attenuator allows to vary the pulse intensity sent to the detector. It is connected to a custom SQS Vl\'aknov\'a Optika 1$\times$400 splitter which distributes the light to every bar of BAND. 

Each bar has an optical fiber glued to its center with UV-curable glue (see
Fig.~\ref{fig:pic-fiberonbar}). We used an exposed fiber for the fiber-scintillant connection since it was a stable, reliable and easy option. The other end of the fiber is connected to the splitter through a patch panel attached to the BAND frame. 

The photodiode provides reference time for the laser system. The output
signal of the photodiode is inverted and shaped before it is sent to
the BAND read-out system. The photodiode signal is then digitized by the same ADCs and
TDCs which are used for the PMT signals.

\subsection{Electronics and DAQ}

\begin{figure}[tb]
	\centering
	\includegraphics[width=0.48\textwidth]{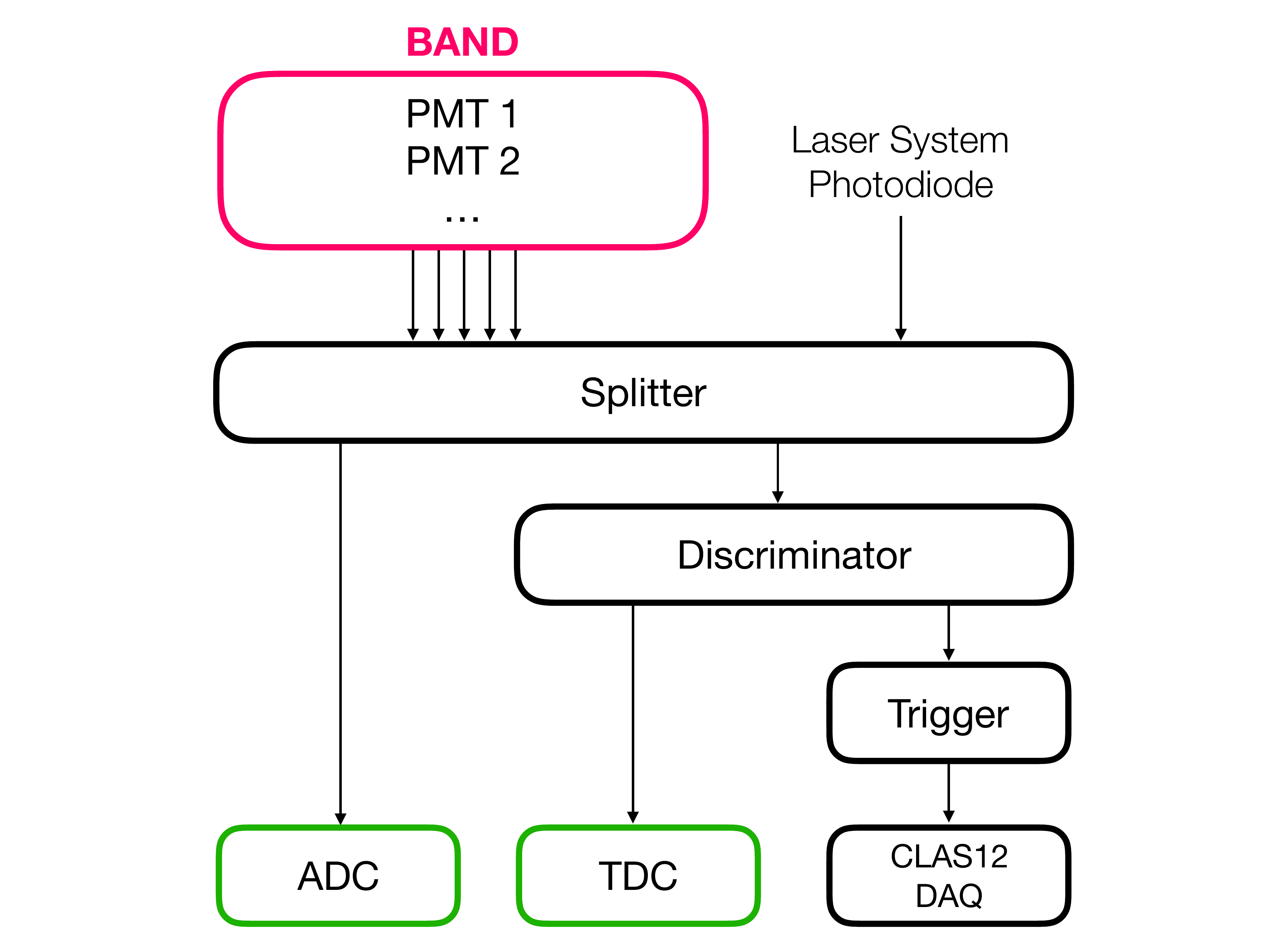}
	\caption{Electronic schematic of the BAND read-out. Every PMT signal is split and ADC and TDC information are obtained. Additional outputs of the discriminators are used to create single bar and cosmic-laser triggers from BAND.}
	\label{fig:electronic-diag}
\end{figure}

The high voltages for each PMT are provided by a multi-channel CAEN SYS4527 mainframe with eleven A1535SN cards with 24 channel each~\cite{caen-hvframe,caen-hvcard}.

A schematic drawing of the read-out electronics and its components is
shown in Fig.~\ref{fig:electronic-diag}. The signal of each PMT is
split to an ADC and TDC, read out independently.  The signal splitters
are custom made and have been used in previous experiments at
Jefferson Lab.  From the splitter one signal is sent to a 250-\si{\mega\hertz} sampling flash-ADCs~\cite{fadc-manual} (FADC) while the
other signal is sent to discriminators.  The discriminated time signal
goes to a TDC (CAEN VX1190A~\cite{caen-tdc}) with 100-ps resolution
per channel.

In total, the system consists of 16 flash-ADCs in one VXS crate, 16
discriminators and two TDCs in a VME crate and 16 splitters.
A trigger signal distribution card for the flash-ADCs and
trigger interface boards are installed in the crates. All components
are part of the standard CLAS12 electronics~\cite{clas12-daq, clas12-trigger}.

The detector is read out either by a trigger from the main CLAS12
trigger system~\cite{clas12-trigger} or by stand-alone BAND 
triggers. The stand-alone triggers are used for tests and calibrations 
with cosmic rays, radioactive sources and the laser system. 
They are implemented by a programmable CAEN V1495 logic board~\cite{caen-logicboard}. Such triggers include
a single bar trigger for source measurements and a
coincidence bar trigger for cosmics and laser measurements. The
coincidence trigger is also fed to the central CLAS12 trigger system
to allow monitoring of the detector by recording laser data during
experimental data taking. This laser-monitoring trigger rate is usually about 10
\si{\hertz}, compared to the $\sim 15-20\,\si{\kilo\hertz}$ trigger
rate from electron interactions in the target.

\subsection{Final Assembly}
\label{sec:assembly}
\begin{figure*}[tbh]
	\centering
	\includegraphics[width=0.48\textwidth , height=0.40\textwidth]{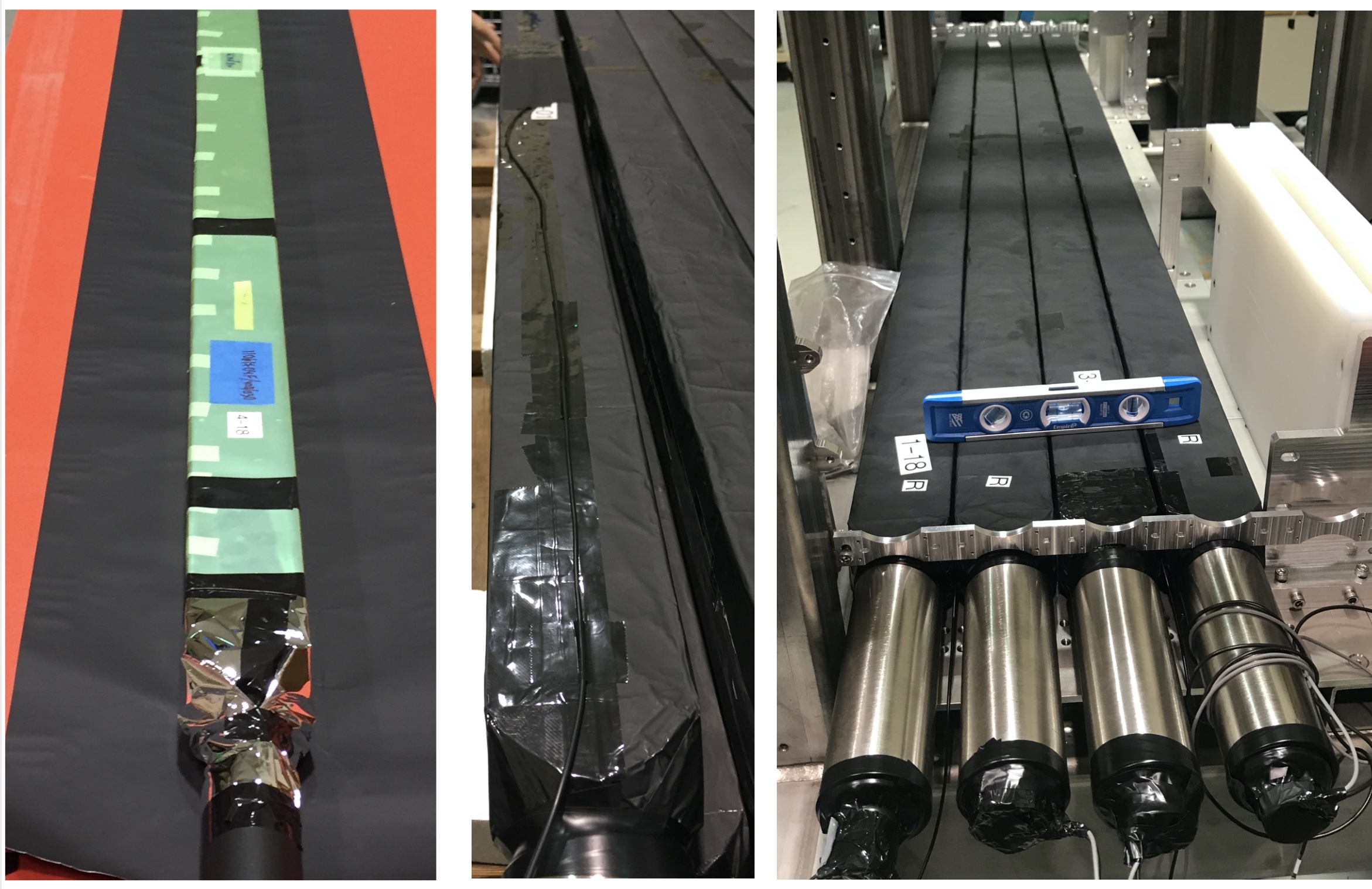}
	\includegraphics[width=0.40\textwidth , height=0.40\textwidth]{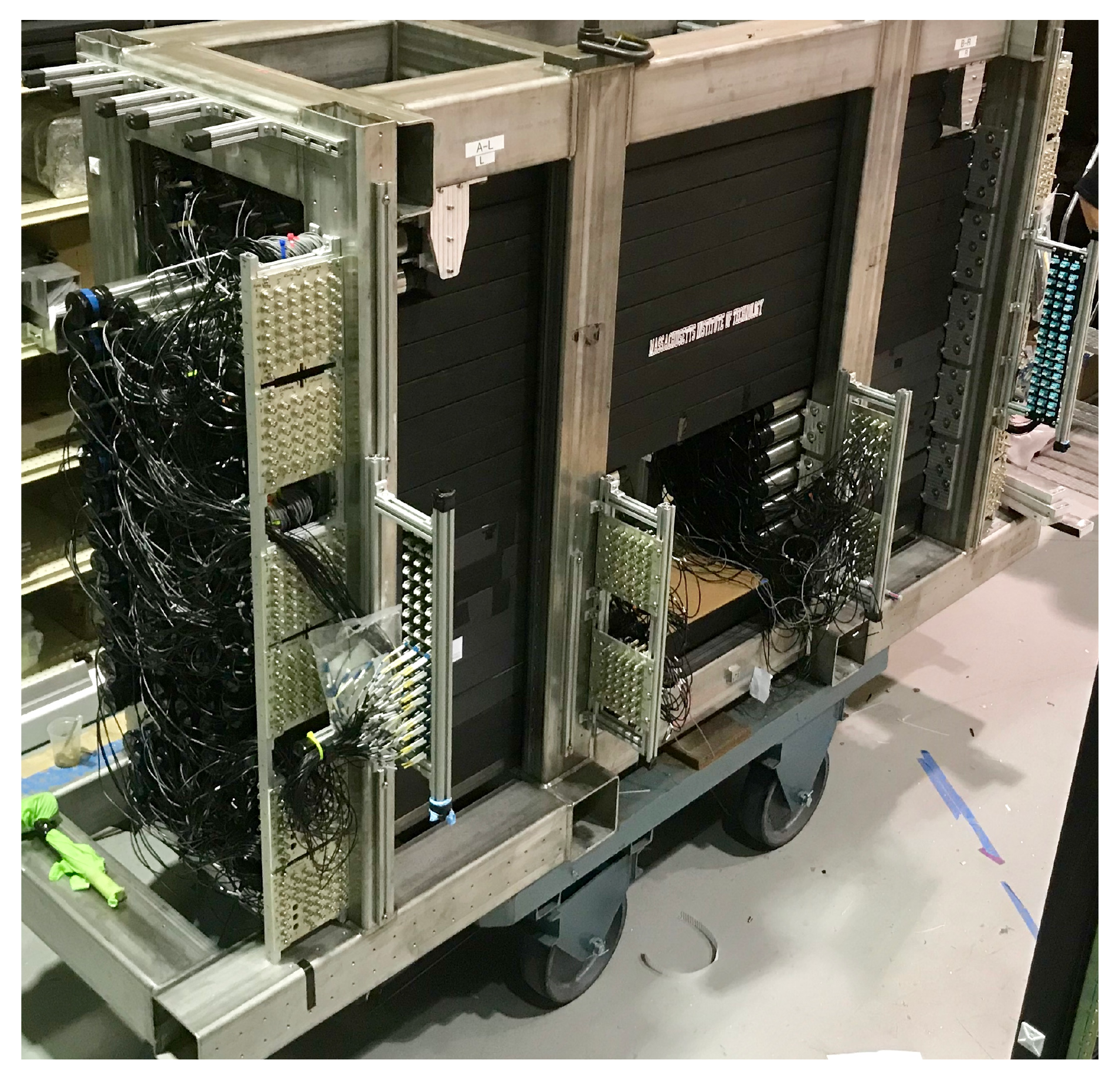}
				\caption{BAND construction. From ESR~\cite{3MESR} wrapped bars (left) to assembled detector (right). The intermediate steps show the fully wrapped bars with the optical fiber and the assembled bottom row of the detector.}
		\label{fig:barassembly}
\end{figure*}

Each scintillator bar was assembled and individually tested, following this procedure:
\begin{enumerate}
\item Select a pair of matched PMTs.
\item Wrap the assembled scintillator with ESR foil~\cite{3MESR}, leaving a window-flap for the laser-system optical fiber (see Fig.~\ref{fig:pic-fiberonbar}).
\item Glue the PMTs to the light guides with MOMENTIVE RTV615 silicone rubber compound~\cite{softglue}.
\item Glue one light guide/PMT assembly to each end of the bar with DYMAX
  UV curing glue~\cite{uvglue}.
\item Wrap with light-tight foil\footnote{Black 50-$\mu$m thick Tedlar\textregistered{} foil~\cite{tedlarfoil} 
was used to wrap the short bars. The long bars were wrapped with Thorlabs black Aluminium foil~\cite{thorlabsfoil}}, leaving a
  window-flap for attaching the laser-system optical fiber.
\item Glue the optical fiber to the center of the bar and reseal the foil window.
\item Install the mu-metal shields on the PMTs.
\item Search for and fix light leaks by measuring the PMT dark currents with a pico-ammeter.
\item Install each bar in the BAND support frame.
\item Connect the optical fibers and signal and HV cables to patch panels mounted on the upstream side of BAND.
\item Final test for light leaks.
\end{enumerate}
Fig.~\ref{fig:barassembly} shows some of the steps in the assembly process,
including one bar wrapped with ESR foil, a fully wrapped bar with the
optical fiber shown, a row of four bars installed in the BAND frame,
and the fully assembled BAND seen from the upstream side with the
patch panels for HV, signal cables, and optical fibers.

\begin{figure*}[tb]
	\centering
	\includegraphics[width=0.48\textwidth]{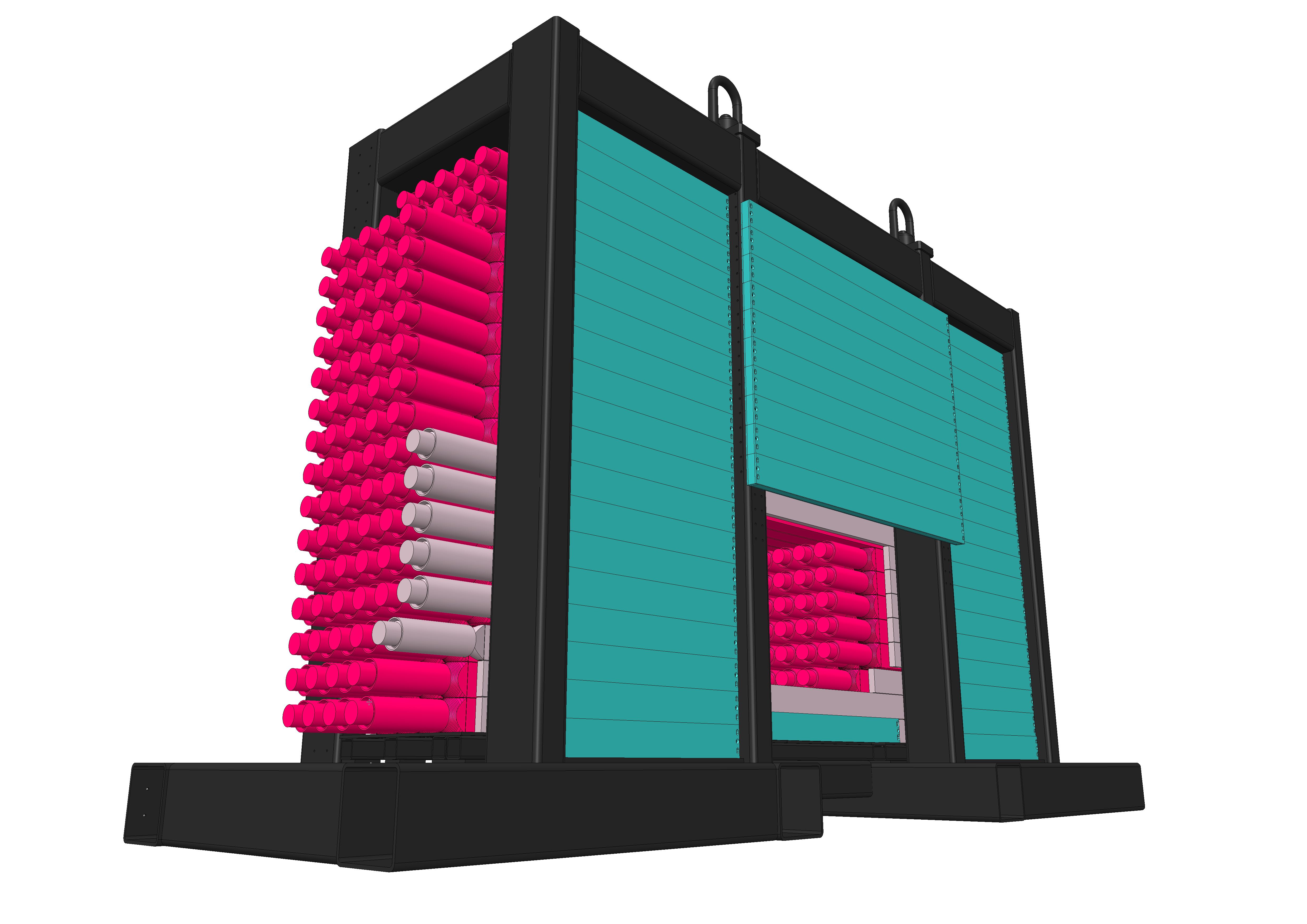} 
	\includegraphics[width=0.40\textwidth]{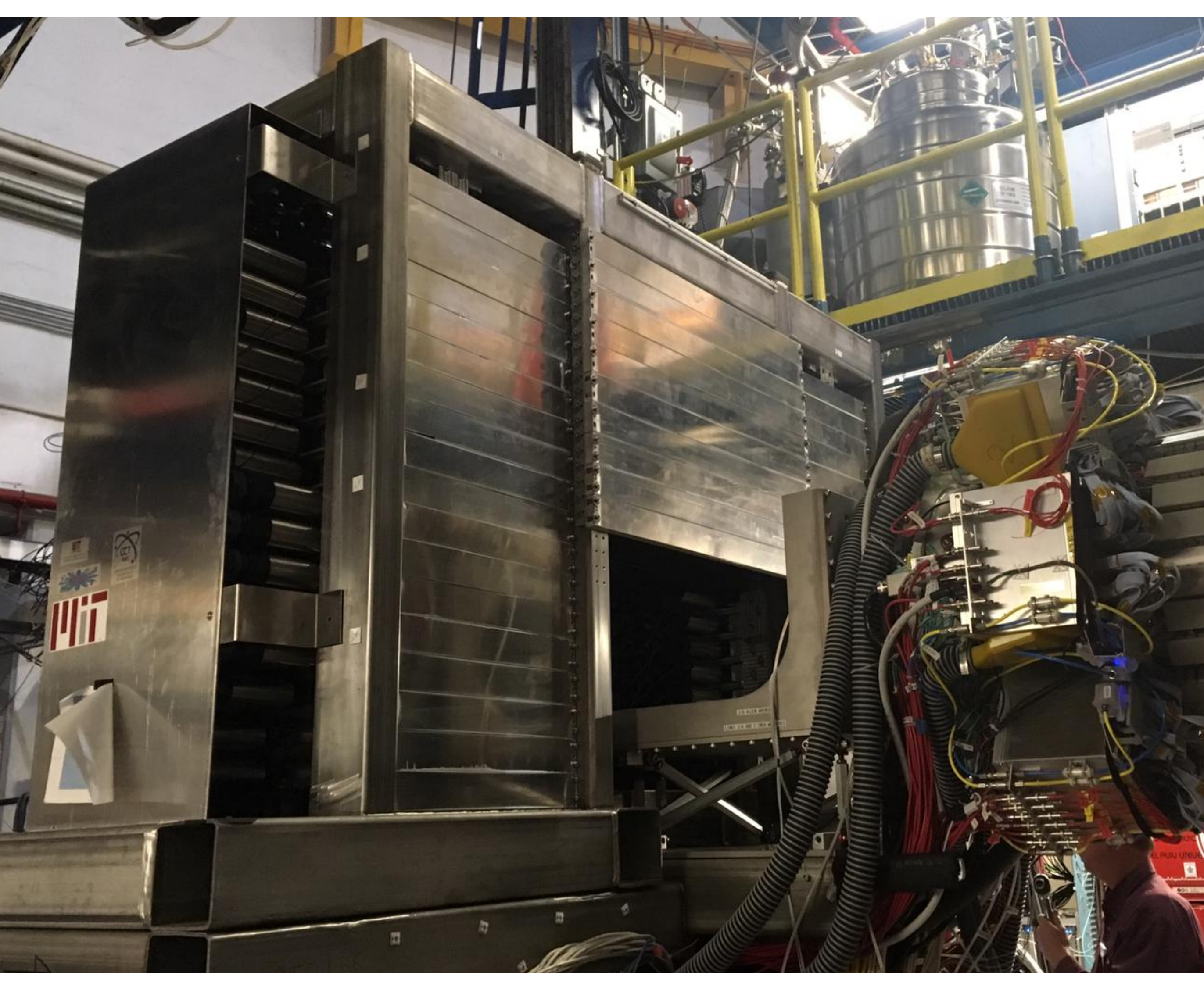}
		\caption{(left) CAD drawing of the downstream side of BAND and its frame. The lead wall is shown in cyan, the scintillators
          are in magenta, and the support frame is in black. (right) Photograph of the downstream side of BAND installed in Hall B. The reflective surface in the photograph are the aluminum covers of the lead blocks.}
		\label{fig:band_downstream}
\end{figure*}

After assembling the bars in the support frame, the lead wall was
installed on the downstream side of BAND. BAND was then
craned into its position on top of the central vertex tracker support
cart upstream of CLAS12. Fig.~\ref{fig:band_downstream} shows a
design drawing and a photograph of the upstream side of BAND. 

\begin{figure}[tb]
	\centering
	\includegraphics[width=0.48\textwidth]{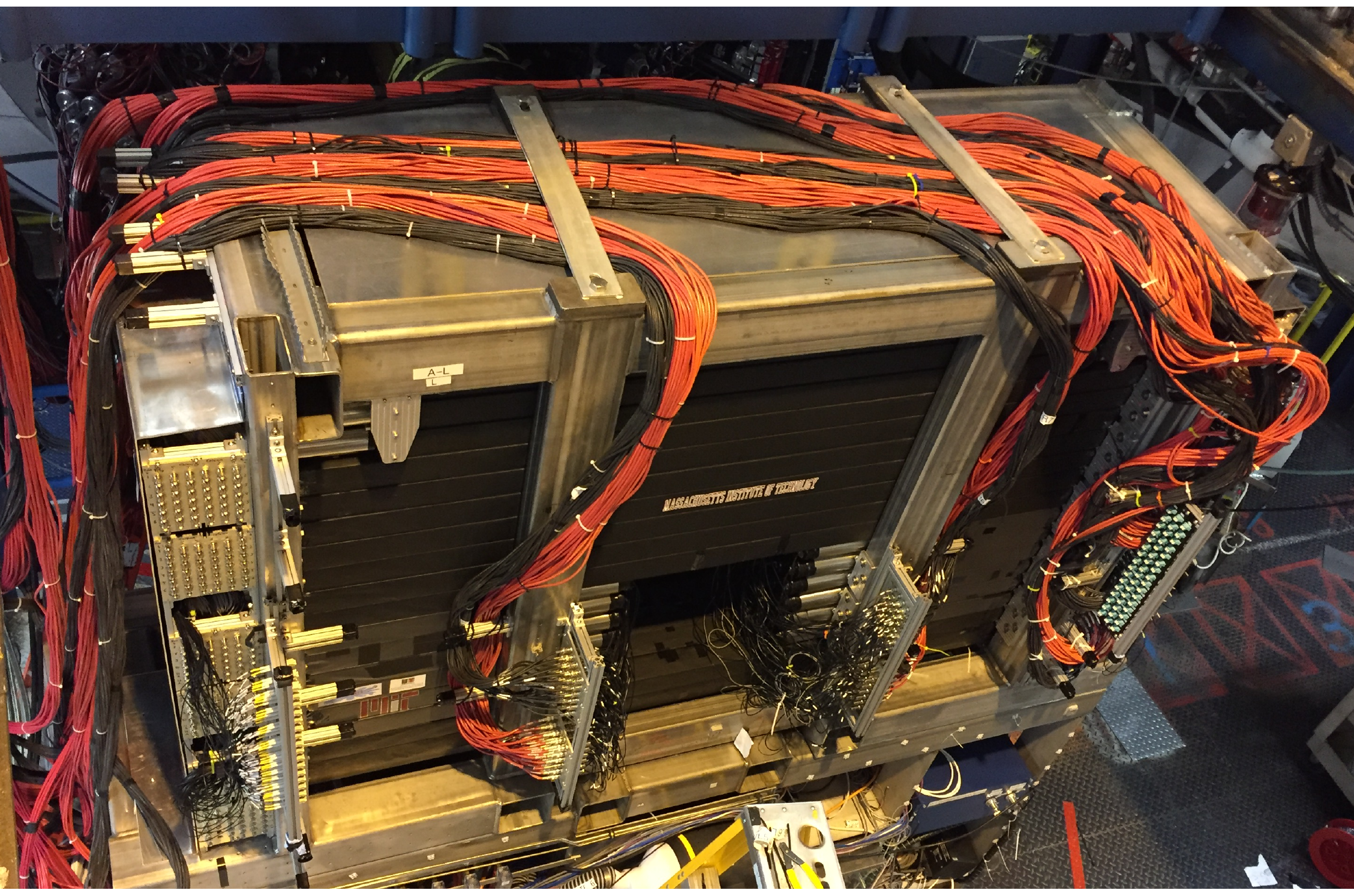}
				\caption{BAND in Hall B. Upstream side with first set of cables installed.}
		\label{fig:bandinhall}
\end{figure}

Fig.~\ref{fig:bandinhall} shows the upstream side of BAND in Hall B with about 3/4 of the cables installed. All cables are connected to
electronics which is installed outside of the photograph on the left side.

\section{Performance}
In this section, we first describe the performance of the individual scintillator bars and their calibrations. We next describe the performance
of the BAND as a whole, with a focus on neutron identification and efficiency.

\subsection{Individual bar performance}
\subsubsection{Gain matching}
\begin{figure}[tb]
	\centering
			\includegraphics[width=0.46\textwidth]{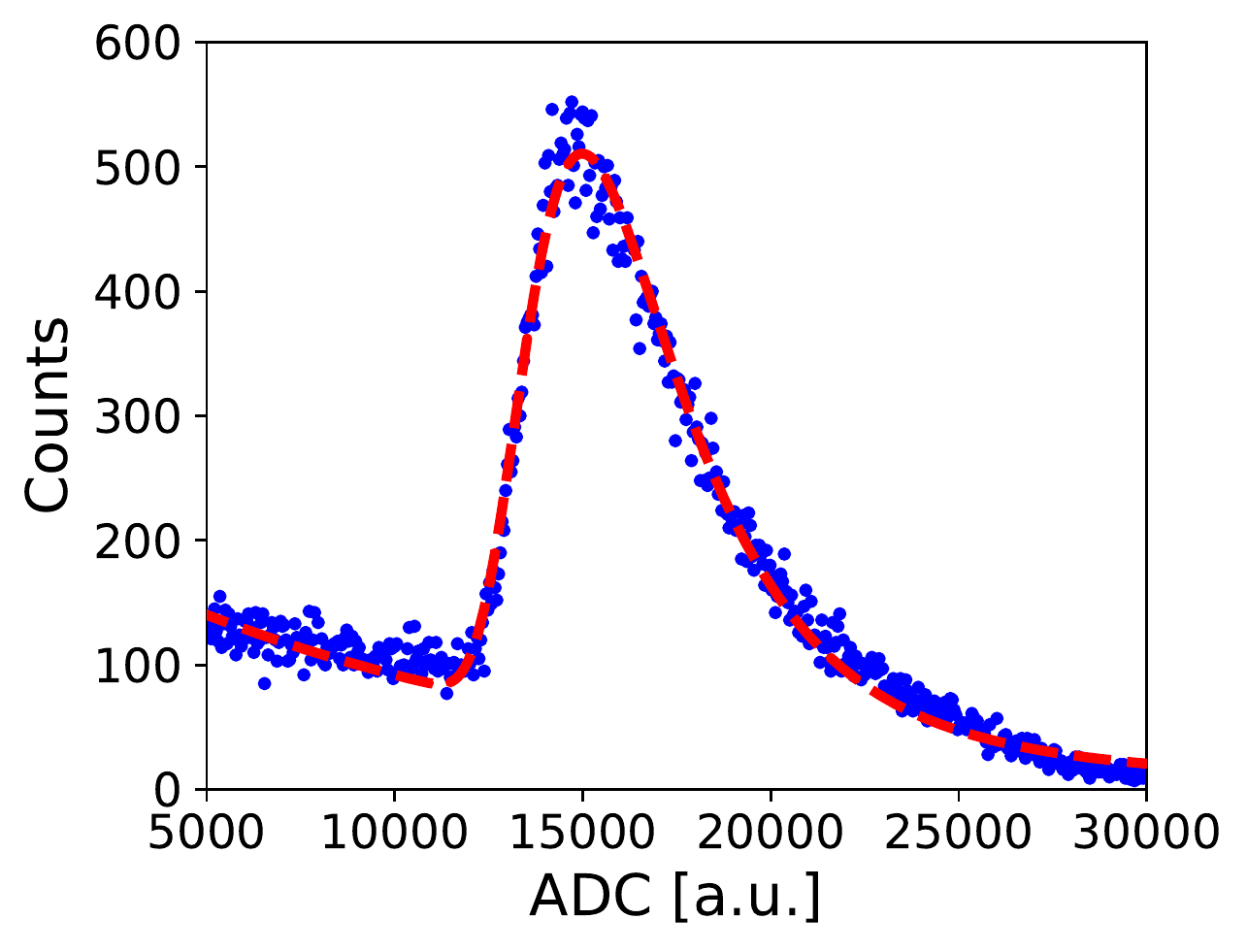}
			\includegraphics[width=0.46\textwidth]{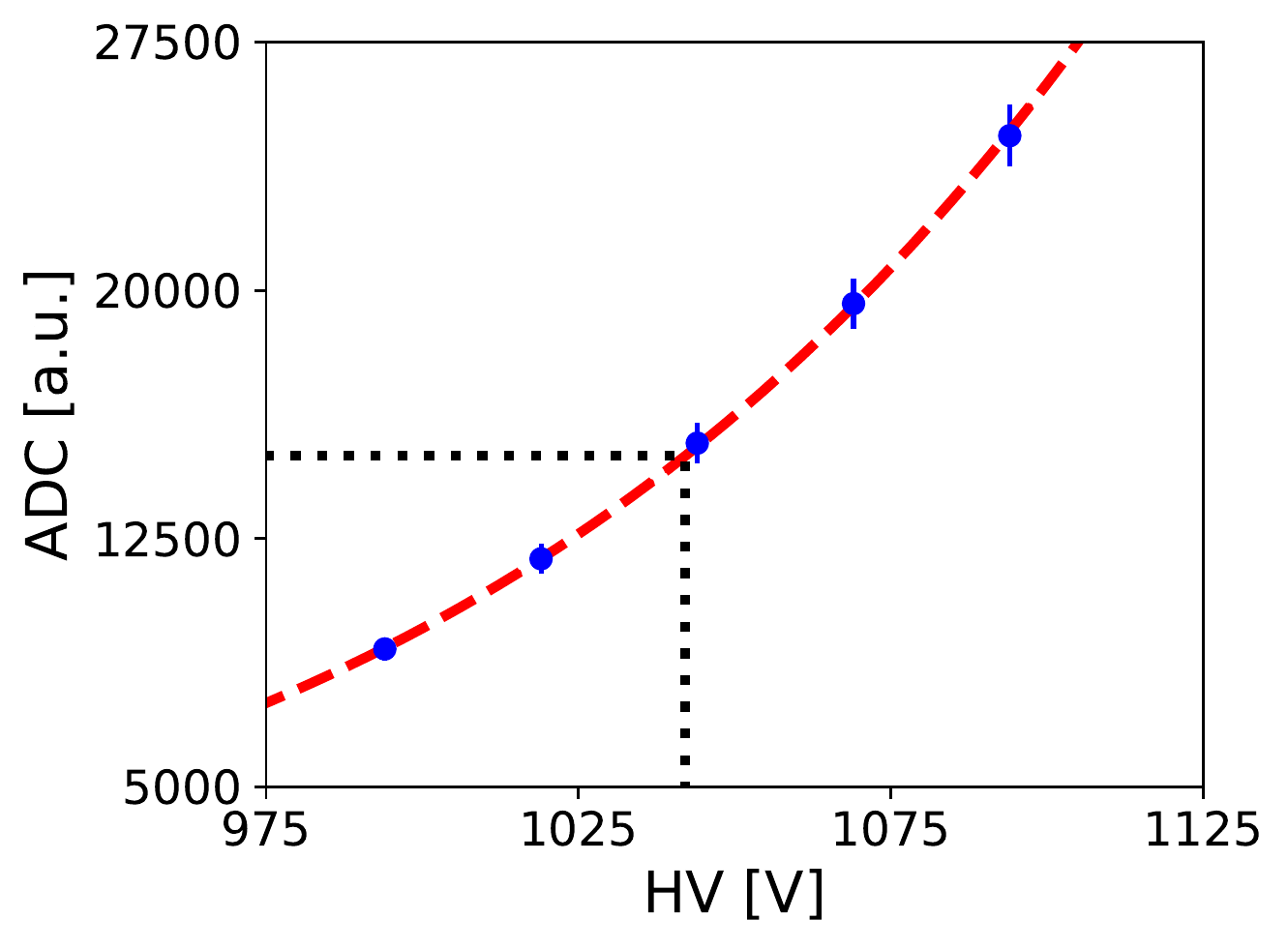}
		
		\caption{Gain curve measurement. (Top) Typical ADC spectrum for a representative PMT with cosmic rays
                  and the fit with a Landau distribution and an exponential background. (Bottom) Gain curve for the same
                  PMT. The dotted lines indicate the final HV setting to set the cosmic peak to ADC channel 15000. }
\label{fig:gain-curve}
\end{figure}

In order to have a similar response to a given energy deposit across all PMTs of BAND, each PMT's HV was optimized using cosmic rays. Cosmic
ray spectra were measured with a range of HV and then combined to produce a gain curve for each PMT. These data were
collected with a cosmic trigger which required a cosmic ray passing through a single vertical layer of BAND to select nearly-vertical
cosmic rays. This corresponds to a relatively high energy deposition of 14.2 \si{\mega\eV}. For each HV setting we fit the ADC spectrum
with a Landau distribution and an exponential background to obtain the cosmic peak position. A representative ADC spectrum is shown in
Fig.~\ref{fig:gain-curve} (top).  The obtained gain curve for this PMT is shown in the bottom panel of Fig.~\ref{fig:gain-curve}. The dashed lines indicate the final HV
setting to position the cosmic peak at ADC channel 15000. This desired peak
 position of cosmic rays was chosen to avoid signal overflows in the sampling FADCs for neutron-induced signals, which deposit less energy as compared to cosmic rays. These overflows dominate above ADCs of 20000. 

\begin{figure}[tb]
	\centering
			\includegraphics[width=0.46\textwidth]{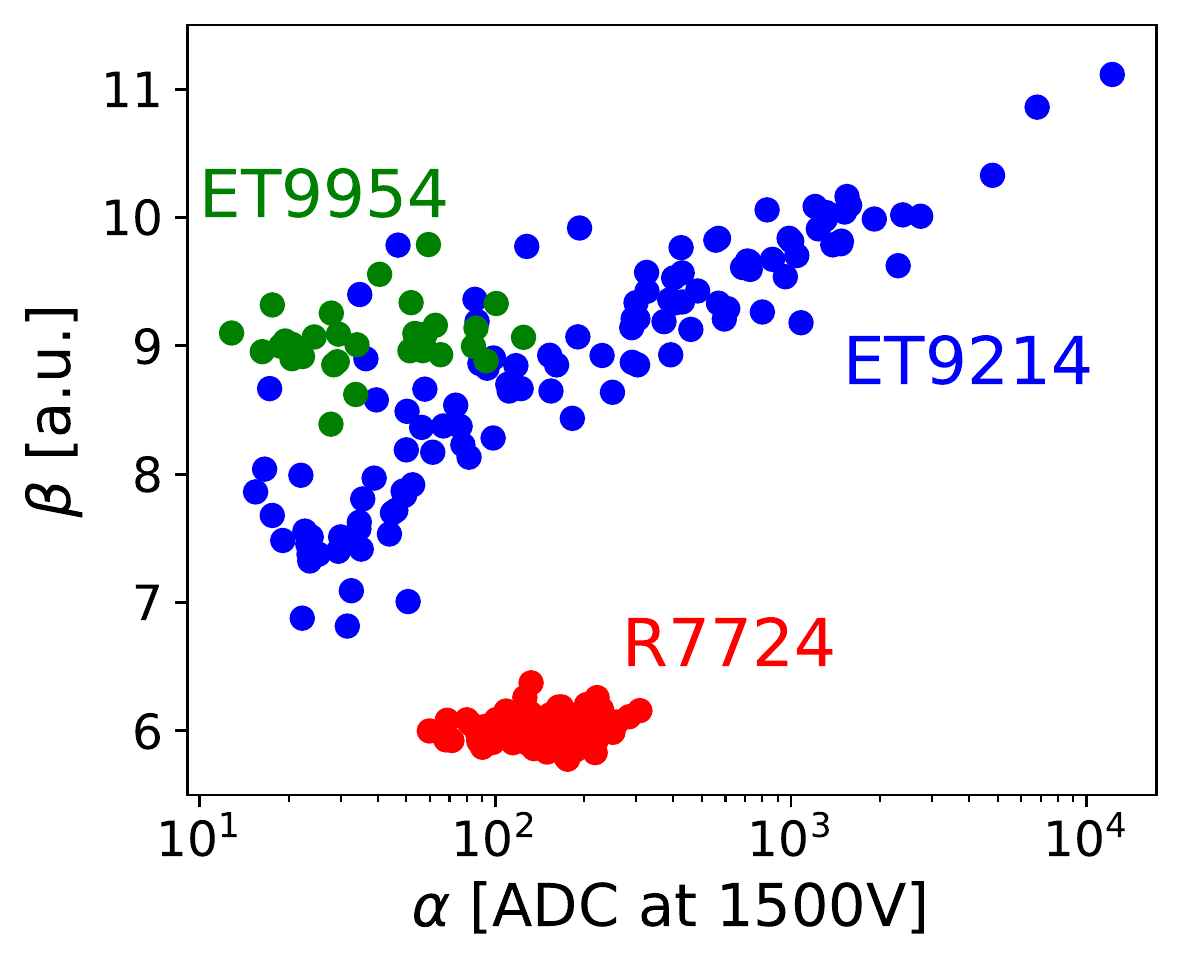}
			\includegraphics[width=0.46\textwidth]{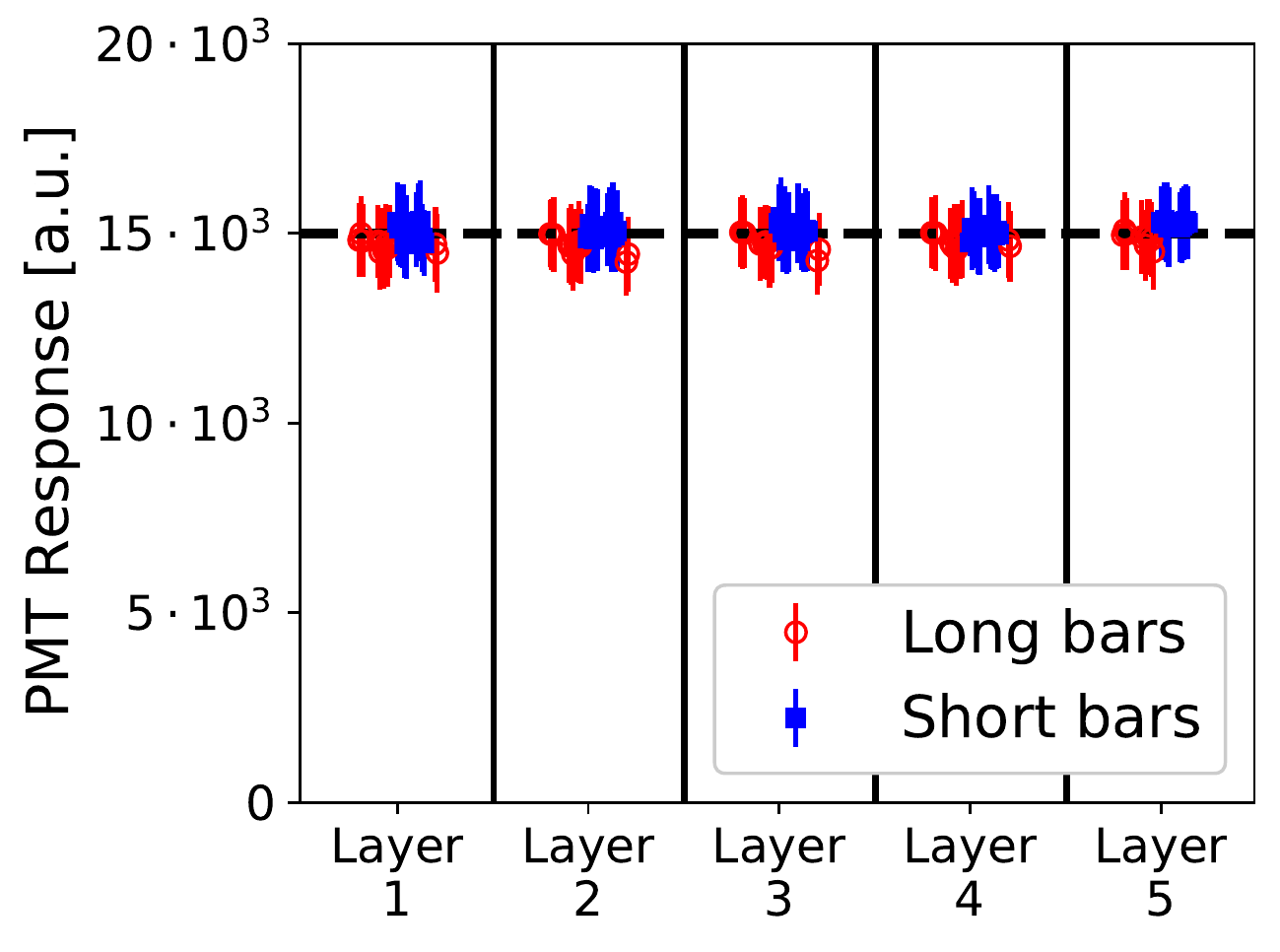}
		
		\caption{ (Top) Gain parameters for each PMT. One sees a greater spread in the Electron Tube PMTs (green and blue) than in the Hamamatsu PMTs (red). (Bottom) $\mathrm{ADC}_{\mathrm{cosmic peak}}$ position for each bar (vetos not shown) after gain matching. All bars are well aligned.}
		\label{fig:hv_settings}
\end{figure}

Each PMT gain curve was fit with a power law with parameters $\alpha$ and $\beta$
\begin{eqnarray}
	ADC	= \alpha \cdot \left(\frac{\mathrm{HV}}{1500}\right)^{\beta},				
		\label{eqn:gain_curve}
\end{eqnarray}
where the division by 1500 \si{\volt} was an arbitrary normalization to improve fit convergence. The value of the fit parameters for each PMT are
shown in Fig.~\ref{fig:gain} (top). The Hamamatsu R7724 PMTs (red) are more uniform than either of the two ET PMTs, 9954KB (green) or
9214KB (blue)\footnote{We obtained similar results when using the pulse amplitude of the ADC signal instead of the integral}. 
However, the cosmic ray ADC peaks are well aligned after adjusting the HV based on the obtained gain curves (see bottom panel in Fig.~\ref{fig:hv_settings}).


\subsubsection{Energy deposit calibration}
\label{sec:energydeposit}
\begin{figure}[tbh!]
	\centering
		\includegraphics[width=0.48\textwidth]{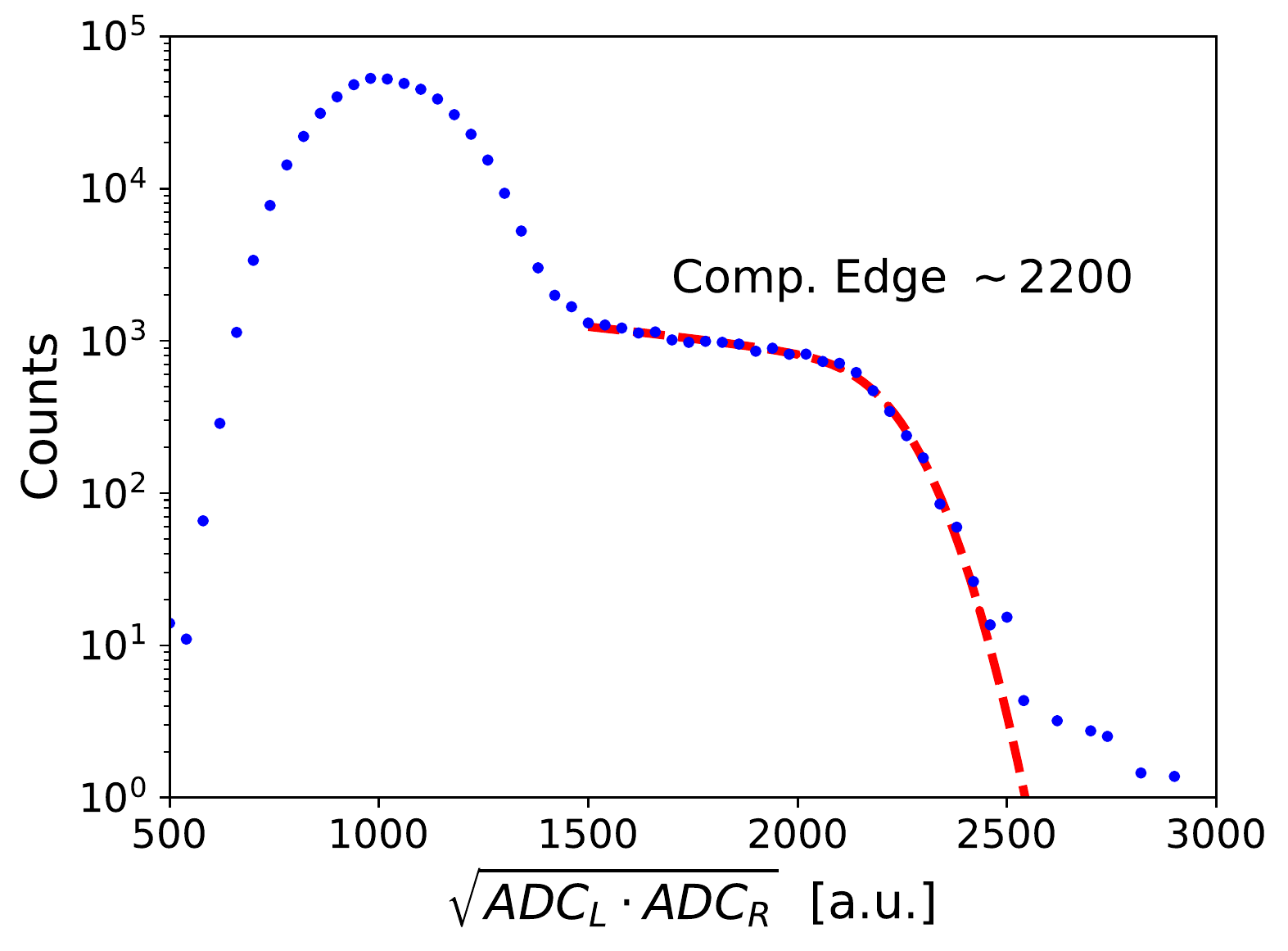}
		\caption{ADC spectrum for a $^{60}$Co placed on the center of the bar along with the fit of the Compton edge.}
	\label{fig:compton_edge}
\end{figure}

The ADC response of each PMT is converted to \si{\mega\electronvolt}
energy deposition by measuring the response to multiple radioactive
sources and to cosmic rays. The sources $^{60}$Co, $^{22}$Na, and $^{137}$Cs were chosen due
to the gamma rays that have Compton edges of $0.963$ and $1.118$,
$0.477$, and, $1.062$ and $0.341$, respectively (in
\si{\mega\electronvolt}). The response to each source was measured in
the center of the bar. Using the ADC response to all sources, in
combination with the attenuation length of the bar, the ADC response
can be converted to \si{\mega\electronvolt}. These measurements were
done for a subset of the BAND bars (ones that were accessible following
installation) after the bars were gain matched (see previous section).

A typical response of a PMT to $^{60}$Co is shown in
Fig.~\ref{fig:compton_edge} along with a fit of the Compton edge by a
parametrization described in~\cite{comptonedge}. The extracted Compton edges
for various bars are quite similar which gives us confidence in
applying these measurements to all bars. 

\subsubsection{Time-walk calibration}
Time-walk calibrations were performed for each PMT with the laser system~\cite{band-laser} by using the fiber optic attenuator~\cite{attenuator} to vary
the amount of light delivered to each bar. Waveforms and times were measured as the attenuator scans from $40$ \si{\decibel} to $0$ 
\si{\decibel}. The photodiode output is used as a reference time for any calibrations 
performed with the laser system, and is external to the variable attenuator.

\begin{figure*}[tb]
	\centering
		\includegraphics[width=0.48\textwidth]{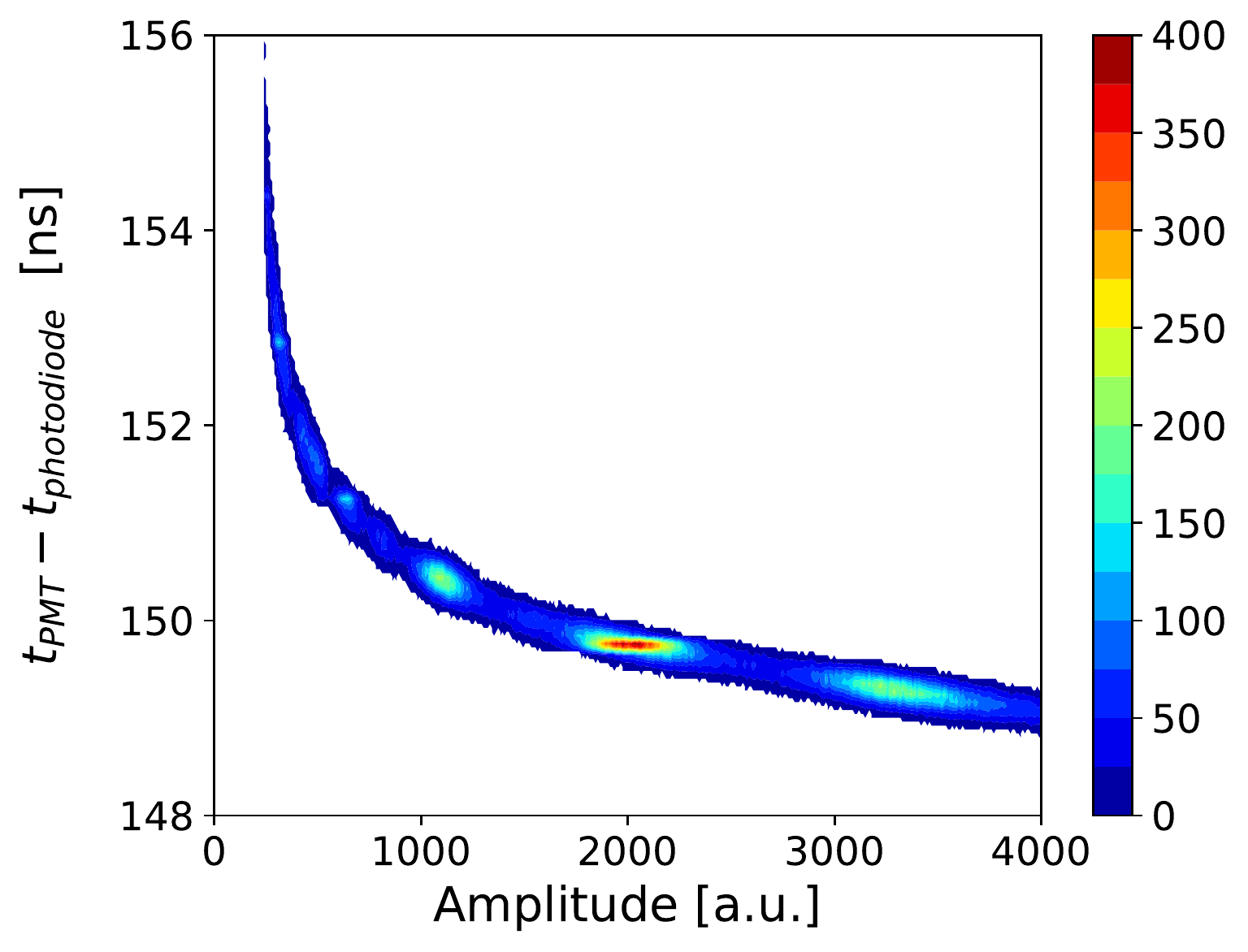}
		\includegraphics[width=0.48\textwidth]{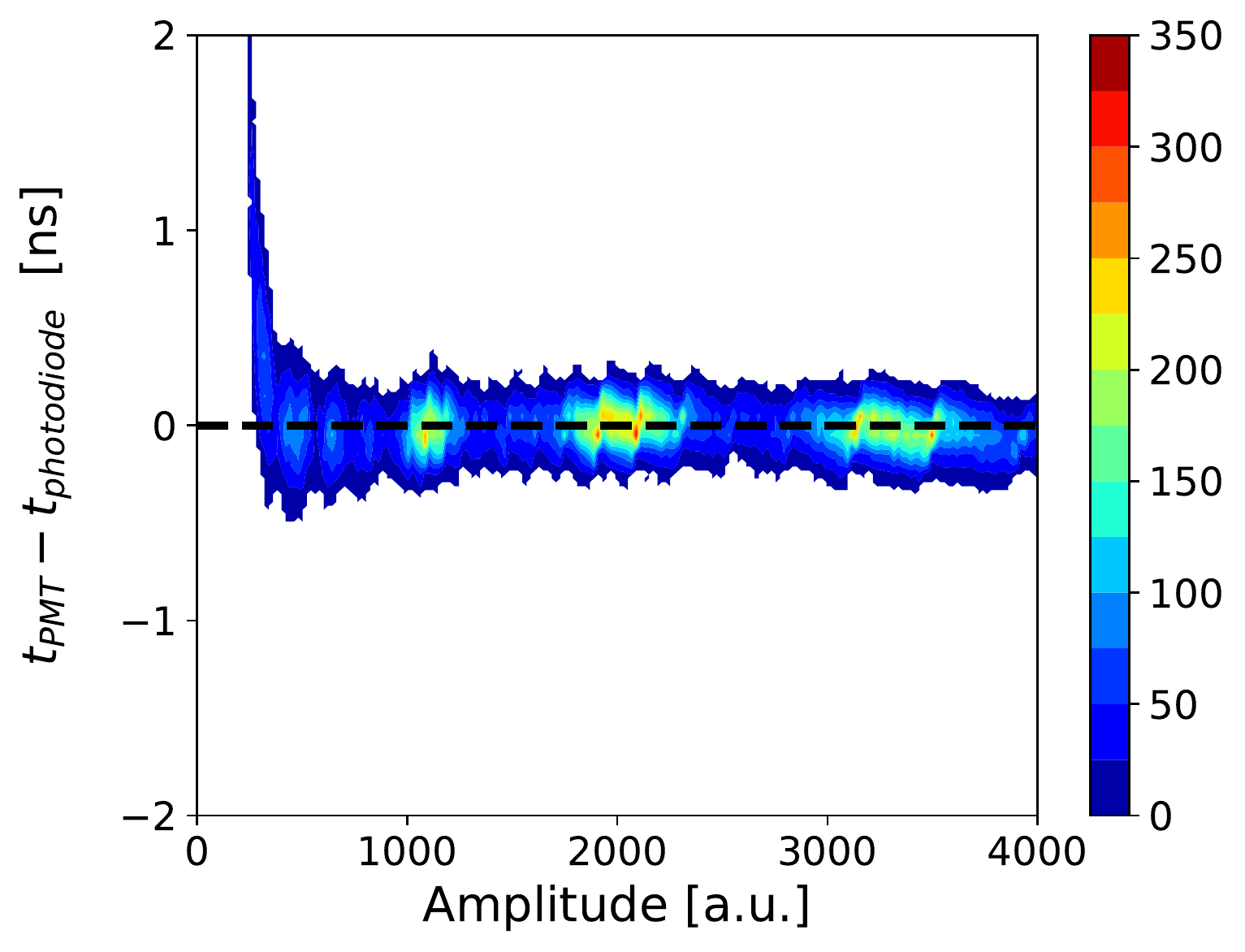}
	\caption{Time-walk calibration: Typical PMT-reference
          photodiode time difference versus pulse height spectrum for one PMT before time-walk correction (left) and after time-walk correction (right).}
	\label{fig:time_walk}
\end{figure*}

Dependence of a typical PMT time on pulse height is seen in
Fig.~\ref{fig:time_walk} (left). To correct for time-walk, we
parameterize this dependence on the pulse height of the ADC signal, $A$,
as:
\begin{eqnarray}
	\begin{split}
		t_{\mathrm{PMT}}-t_{\mathrm{photodiode}}	&= \alpha + \frac{\beta}{\sqrt{\textrm{A}}}.				
		\label{eqn:time_walk}
	\end{split}
\end{eqnarray}
The residual difference, after the time-walk correction,
$t_{\mathrm{PMT}}-t_{\mathrm{photodiode}}$, is shown in
Fig.~\ref{fig:time_walk} (right). At pulse heights close to threshold,
the parameterization is not flexible enough and underestimates the
strong dependence on pulse height. However, signals with pulse heights
this low are not of interest, as we found they are dominated by background,
rather than by neutrons (see Section~\ref{sec:neutronidentification}). Any residual corrections that are needed at higher pulse heights are corrected for iteratively, using the same equation as above. 

\subsubsection{Effective velocity}
\begin{figure*}[tb]
	\centering
		\includegraphics[width=0.47\textwidth]{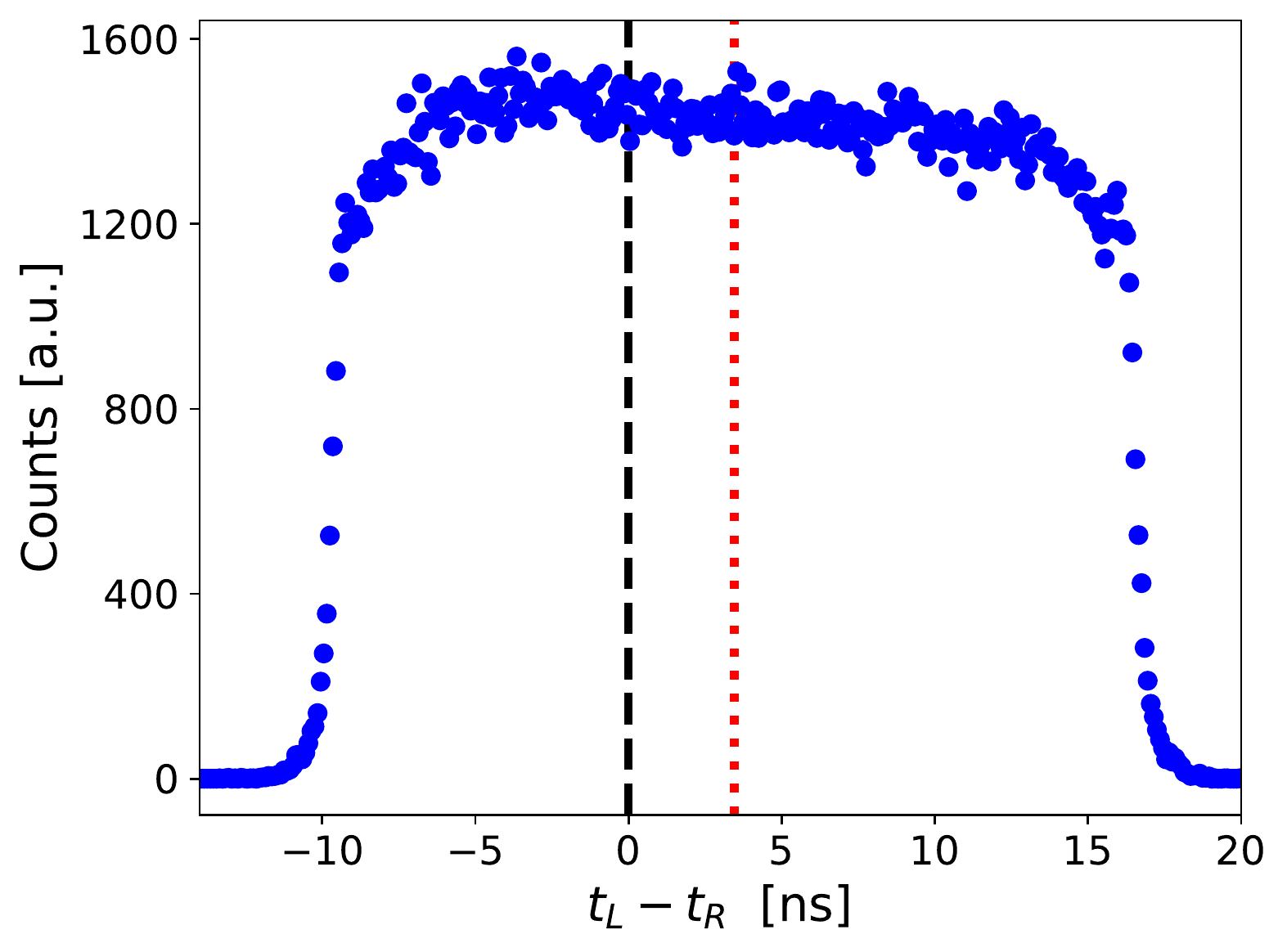}
		\includegraphics[width=0.47\textwidth]{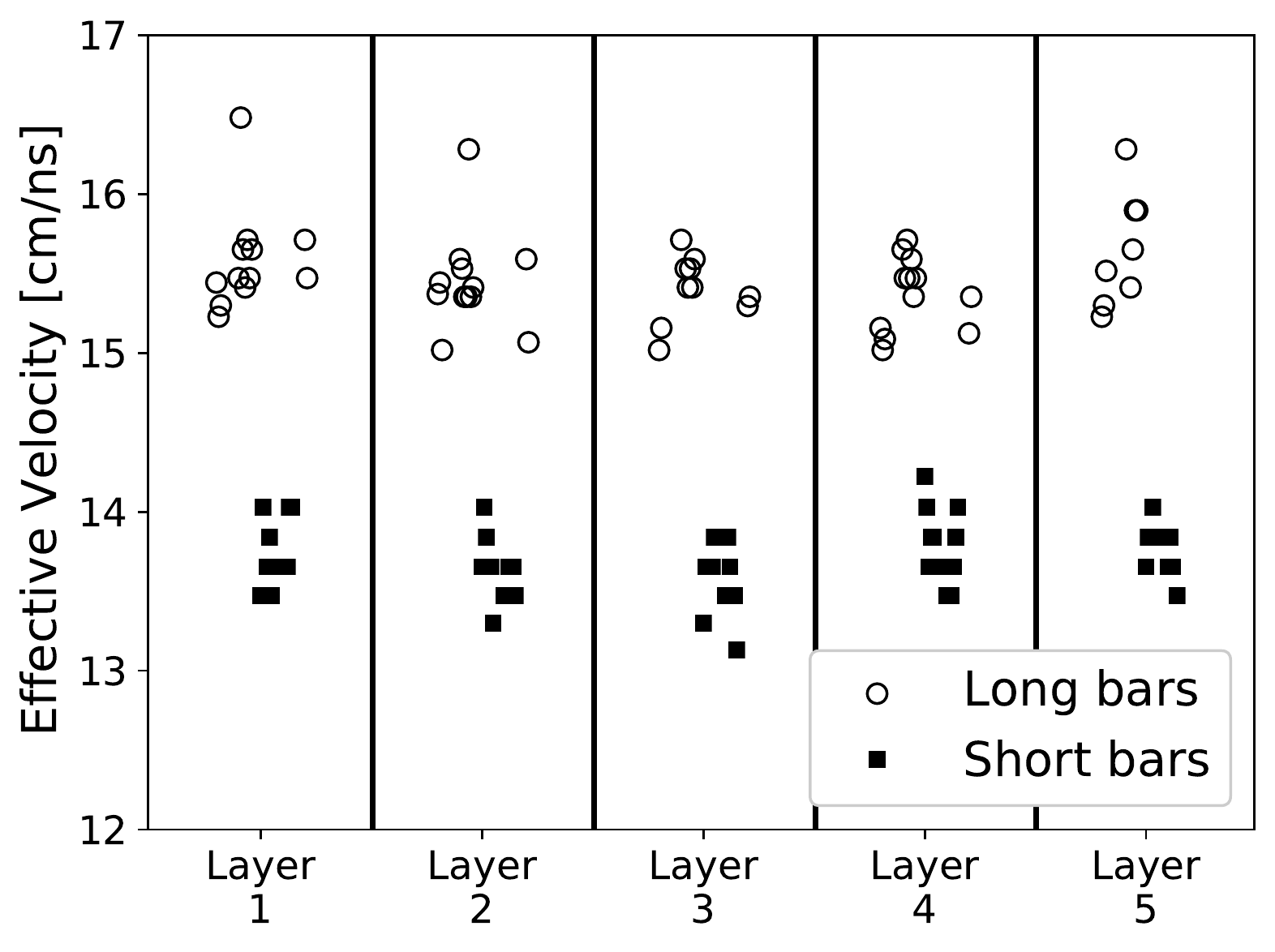}
                \caption{ (Left) Typical time-walk corrected $t_{L} -
                  t_{R}$ time spectrum in a single bar. The red
                  dotted line indicates the $LR$ offset value from
                  0. (Right) Effective light-speed for each
                  bar. The differences in effective speeds between
                  long bars (circles) and short bars (boxes) are due
                  to geometrical effects.}
	\label{fig:eff_vel}
\end{figure*}

The relative time delays between PMTs on the same bar and the
speed-of-light in that bar are extracted using cosmic ray data. 
The effective light speed in
a given bar is extracted from the width of the
relative timing distribution between its left and right PMTs and the bar's physical length:
\begin{eqnarray}
	\begin{split}
		t_L - t_R 	= -\frac{2x}{v}							\\
		 -\frac{L}{v}	\leq 	t_L - t_R 	\leq \frac{L}{v},				
		 \label{eqn:eff_vel}
	\end{split}
\end{eqnarray}
The relative left-right PMT time offset is given by the center of the
distribution (see Fig.~\ref{fig:eff_vel} left).  This procedure is done after the PMT timing information has been time-walk corrected. The obtained effective
velocities for each bar are shown in Fig.~\ref{fig:eff_vel} (right)
distinguished between short bars (boxes) and long bars (circles). The
observed difference is due to geometrical effects from the different
bar lengths.

\subsubsection{Attenuation length \label{attenlen}}
\begin{figure*}[tb]
	\centering
		\includegraphics[width=0.48\textwidth]{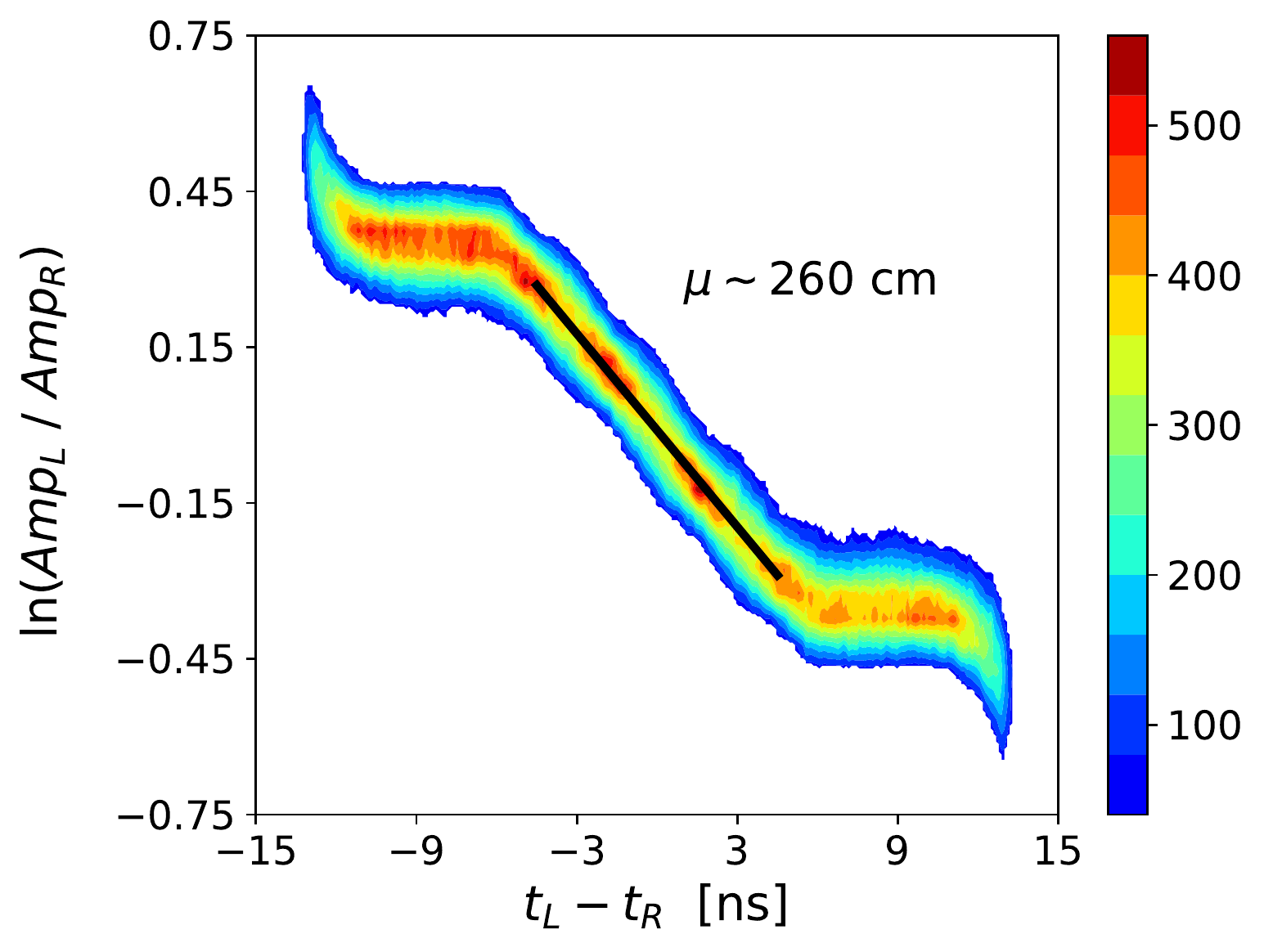}
		\includegraphics[width=0.48\textwidth]{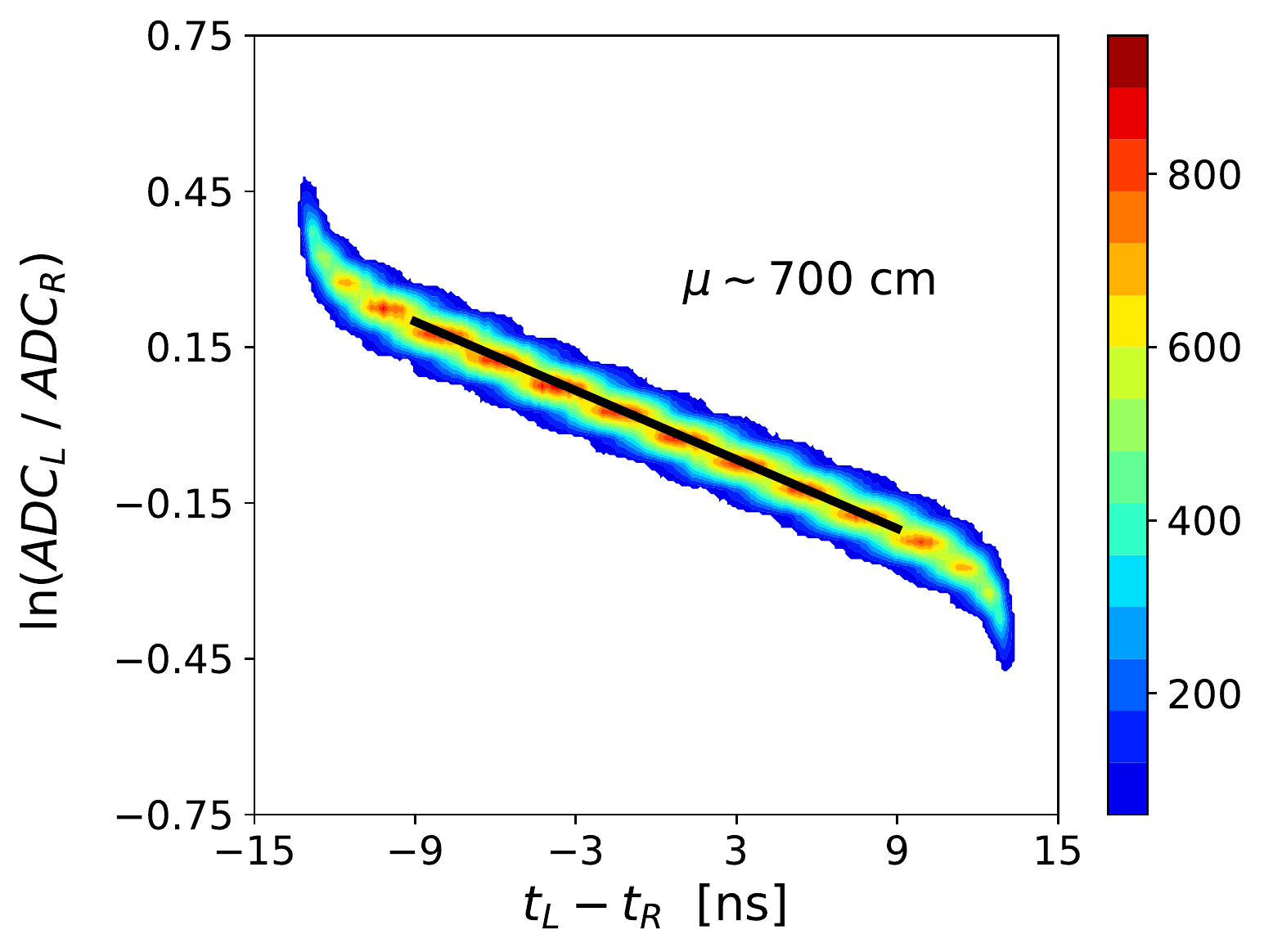}
	\caption{(Left) Log ratio of ADC amplitudes from both PMTs on a bar as a function of time difference of the PMTs. (Right) Log ratio of ADC integral from both PMTs as a function of time difference.}
	\label{fig:atten}
\end{figure*}

We measured the effective attenuation length of each completed
(assembled and wrapped) bar from cosmic ray data by using the ADC
amplitudes ($A_L,A_R$) and times ($t_L,t_R$) measured by the two PMTs:
\begin{eqnarray}
	\begin{split}
		A_L(x) &= A_0 e^{-\frac{1}{\mu}\left(L/2-x\right) }				\\
		A_R(x) &= A_0 e^{-\frac{1}{\mu}\left(L/2+x\right) }				\\
		R(x) \equiv \ln{\frac{A_L(x)}{A_R(x)}} &= \frac{2x}{\mu} = \frac{-v(t_{L}-t_{R})}{\mu}			
		 \label{eqn:atten}
	\end{split}
\end{eqnarray}
where $x$ is the location of the cosmic-ray hit, $\mu$ is the
attenuation length, $A_0$ is the amplitude of cosmic ray interaction,
and $v$ is the effective speed of light in the bar.  A typical plot of
$R(x)$ as a function of $t_{L}-t_{R}$ is shown in Fig.~\ref{fig:atten}
(left).  We fit the slope of the central part to obtain the
attenuation length.  The change in slope of the amplitude ratio closer
to the edges of the bar is caused by light reflections from the light
guides, which affected the amplitude and signal timing for events close to one
PMT. We verified this by seeing the reflections in the unintegrated
Flash ADC spectrum.  We also see this using the integral of the signal (over
80 ns) rather than
its amplitude (its peak) for $R(x)$, see Fig.~\ref{fig:atten} (right). The
distribution is linear over the whole bar since most of
the reflections are included in the large integration window of 80
\si{\nano\s}. However, the attenuation lengths from the integrated
signals are very different from the bulk attenuation lengths and those
measured in the bench tests due to the multiple reflections.

\subsubsection{Individual Bar offsets}
After all bars are individually calibrated, relative time delays between bars can still remain, and require correction to combine 
statistics from all bars. The alignment of all bars can be done quickly using the laser calibration system, as the laser pulse arrival 
time to each bar should be within the timing resolution. Any residual offset 
between bars can be corrected later by using the photon arrival time
at each bar with respect to electrons detected in the CLAS12 forward
detector.

We used the laser calibration system to send light pulses to all the
bars.  We aligned the average time of each bar, 
$t_{avg,i} = \frac{1}{2} \left(t_L + t_R\right)_i$, with respect
to a reference bar in its layer and then we aligned the reference bars
to that of layer 5.  
For example, the alignment for bar $i$ in layer
$j$ is done as follows:
\begin{eqnarray}
	\begin{split}
		\mathcal{O}^{ij} 	&= \braket{ t_{avg}^i - t_{avg}^{j}  }				\\
		\mathcal{O}^{j} 		&= \braket{ t_{avg}^j - t_{avg}^{ref}  }				\\
		t^{i,j}_{corr} 		&=  t_{avg}^i - \mathcal{O}^{ij}  - \mathcal{O}^{j},
		\label{eqn:bar_offsets}
	\end{split}
\end{eqnarray}
where $ t_{avg}^{ref}$ is the average time of the global reference bar chosen on the most downstream layer of BAND, and
$t_{avg}^j$ is the average time of the reference bar chosen in layer $j$. Then only one global offset is needed to correct the
offset of $t_{avg}^j$ if there are no residual offsets between bars.

\subsection{BAND performance} 
\subsubsection{Global offset}
\label{sec:global_offset}
A final global time offset is required to align BAND with
CLAS12. The CLAS12 time is determined by the time of the electron detected in the
forward detector, as traced back to the target.

\begin{figure}[tbh!]
	\centering
		\includegraphics[width=0.48\textwidth]{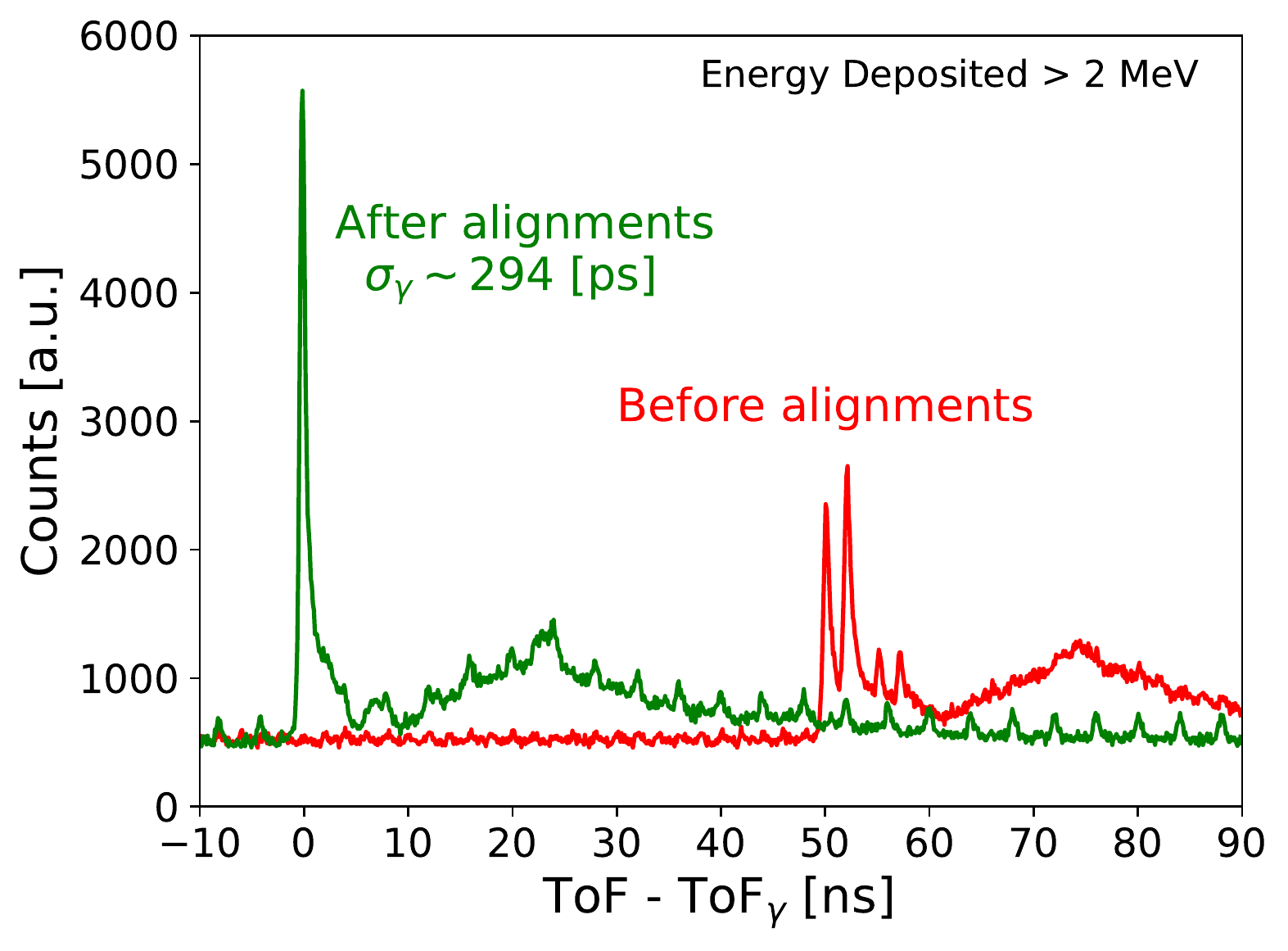}
                \caption{Relative electron-BAND time difference
                  spectrum for all bars offset by the photon travel
                  time for each bar: (red) after time-walk
                  corrections, but not offset corrections, and (green)
                  after all corrections. The bars are well
                  aligned with a width of the photon peak of
                  $\sigma=294$ \si{\pico\s}. A 2-\si{\mega\electronvolt} energy deposition cut is
                  applied to reduce the random background.  The
                  periodic 4-\si{\nano\s} spikes in the remaining
                  spectrum are due to photons correlated with out-of-time
                  electrons from other beam bunches.}
	\label{fig:final_offset}
\end{figure}

Fig.~\ref{fig:final_offset} shows the uncorrected electron-BAND time difference
spectrum offset by the photon travel time for each bar.  There was a 2-\si{\mega\electronvolt} energy deposition cut to reduce the random
coincidence background. The uncorrected spectrum includes only time walk
corrections but no offset corrections. One can see several sharp
photon peaks between 50 and 60 \si{\nano\s}, far away
from the expected zero value. 

After applying the individual offset corrections from the previous
section, we determined a timing offset for each bar from the location
of its photon timing peak.  Applying these offsets results in the
corrected spectrum, which displays a well-aligned photon peak with a
width of $\sigma=294$ \si{\pico\s}, consistent with the design value.
The periodic 4-\si{\nano\s} spikes in the remaining spectrum are due to photons
correlated with out-of-time electrons from other beam bunches.
 
\subsubsection{Time of Flight resolution}
\label{sec:tofresolution}
\begin{figure*}[tbh!]
	\centering
			\includegraphics[width=0.48\textwidth]{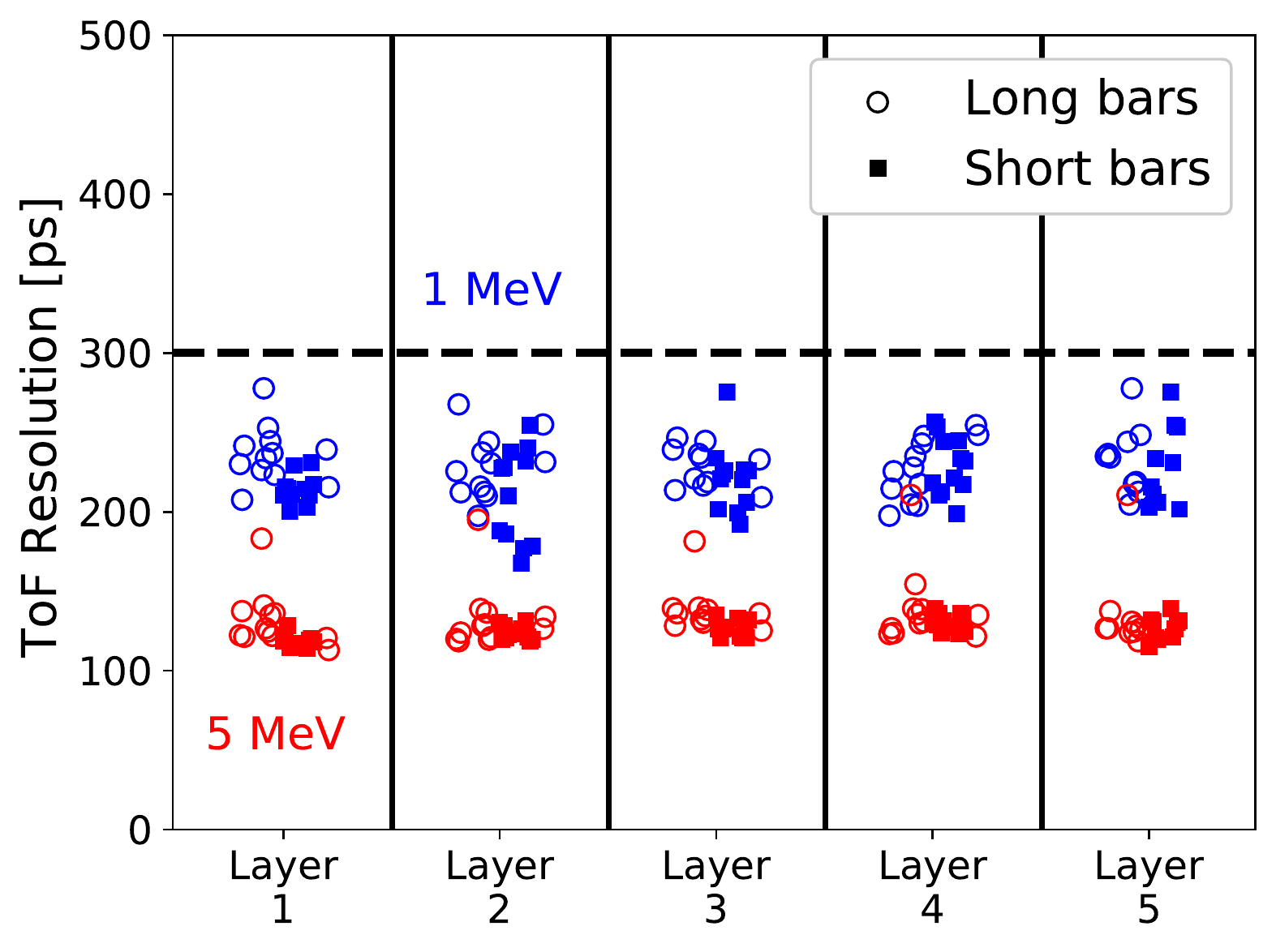}
		\includegraphics[width=0.48\textwidth]{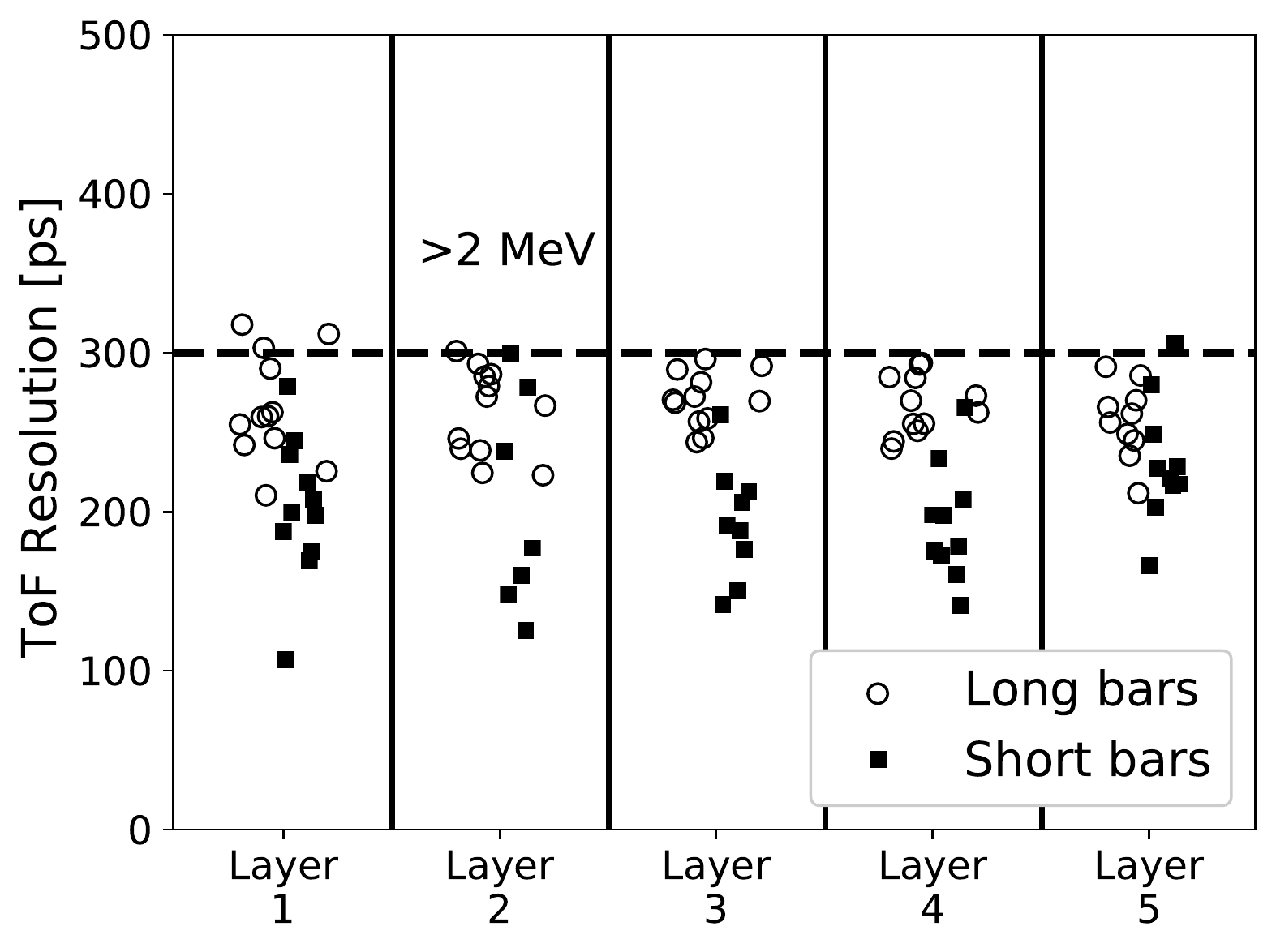}
	\caption{Time resolution for every bar. (Left) as measured
          with the laser calibration system for two separate energy
          deposition cuts. (Right) as measured
          from the electron-photon time difference spectrum with a
          2-\si{\mega\electronvolt} energy deposition cut. All bars
          are at or below the required 300-\si{\pico\s} resolution. }
	\label{fig:tof_resolution}
\end{figure*}

The time resolution of each bar was measured
independently using laser calibration data and using the electron-BAND
time difference spectrum. The laser data allows us to measure the time
resolution over the whole energy deposition range from 0.5 to 7 \si{\mega\electronvolt}  (see
Fig.~\ref{fig:resolution-laser}).

The laser-based resolutions are obtained from a Gaussian fit of the difference of
the average time of a bar to the reference photodiode as a function of energy
deposited, see Fig.~\ref{fig:tof_resolution}. The resolutions for the
1- and 5-\si{\mega\electronvolt} cuts are well below the 300-\si{\pico\s} design goal. For the 5-\si{\mega\electronvolt} cut, which
is a realistic neutron-selection energy deposition cut (see Section
\ref{sec:neutronidentification}) the resolutions are better than 150 \si{\pico\s}.

Fig.~\ref{fig:tof_resolution} (right) shows the time resolutions from
the measured electron-photon arrival time difference with an energy deposition cut of 2 \si{\mega\electronvolt}. We used a high-statistics data run where
electrons are detected in CLAS12. The data was also used for the
global offset corrections presented in Sec~\ref{sec:global_offset}.
These measured resolutions include
additional uncertainties from the pathlength of the target photons
to the bar ($\sigma\approx 70$ \si{\pico\s}) and from the time resolution of the electron beam bunch ($\sigma \approx 25$ \si{\pico\s}). The
laser data does not have these uncertainties.

\subsubsection{Neutron identification}
\label{sec:neutronidentification}
\begin{figure}[tbh!]
	\centering
		\includegraphics[width=0.48\textwidth]{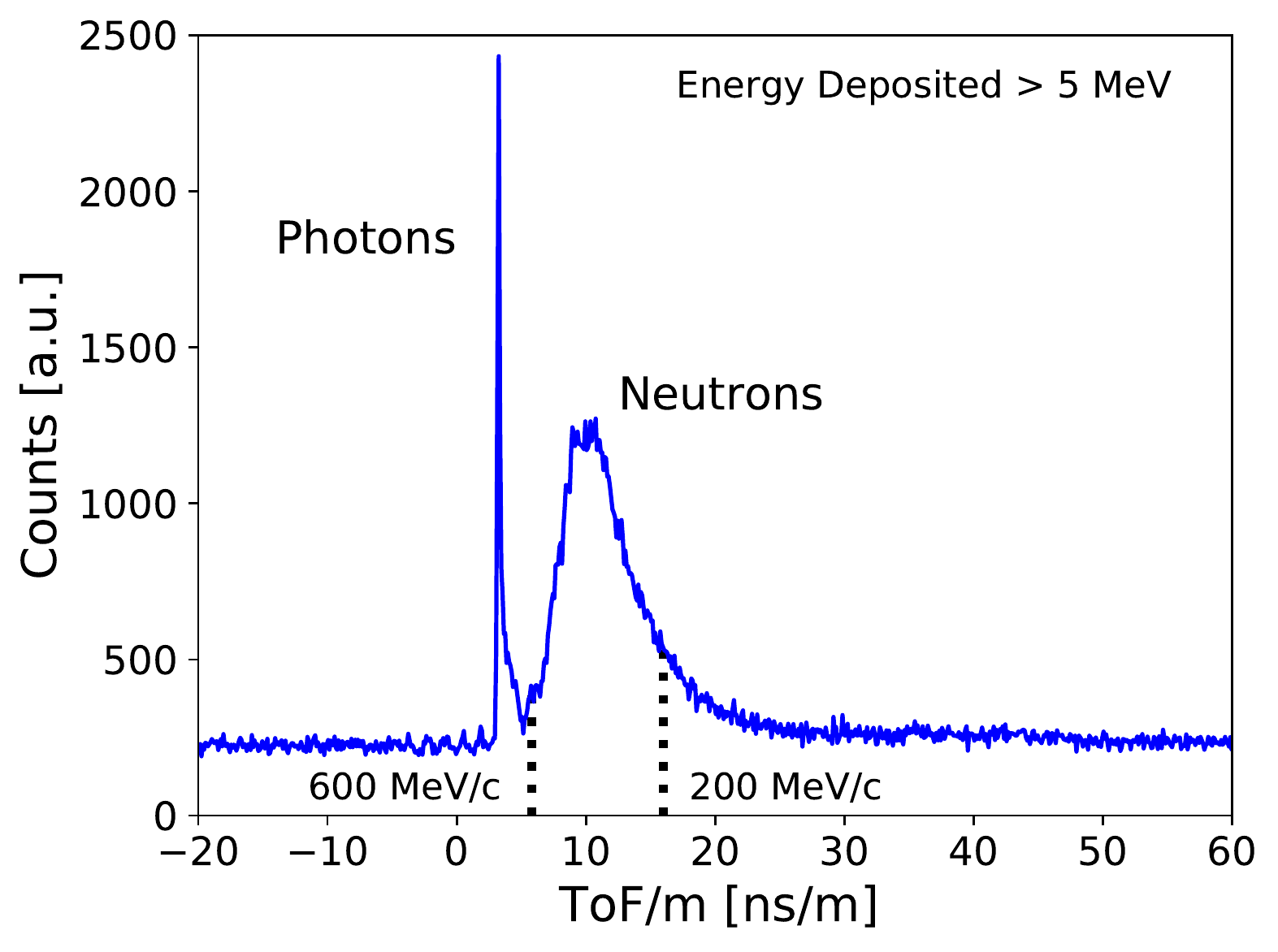}
                \caption{Time-of-flight per meter of BAND hits in
                  coincidence with electrons measured in
                  CLAS12. A sharp photon peak and a broad neutron peak are visible
                  over a flat background from accidentals. The dashed
                  lines indicate the range of neutrons with momenta
                  between $200$ and $600$ \si{\MeV/\clight}.}
	\label{fig:tof}
\end{figure}

We identified neutrons from the electron-deuteron, $d(e,e')X$,
production data by cutting on the
time-of-flight per pathlength (ToF/m), where ToF is the time difference
between the time that the electron
interacted in the target and the particle arrival time measured by
BAND (see Fig.~\ref{fig:tof} for data with a minimum energy deposition
cut of 5 \si{\MeV}). There are well-separated peaks for
photons and for neutrons on top of a flat background from accidental coincidences. 

\begin{figure}[tbh!]
	\centering
		\includegraphics[width=0.48\textwidth]{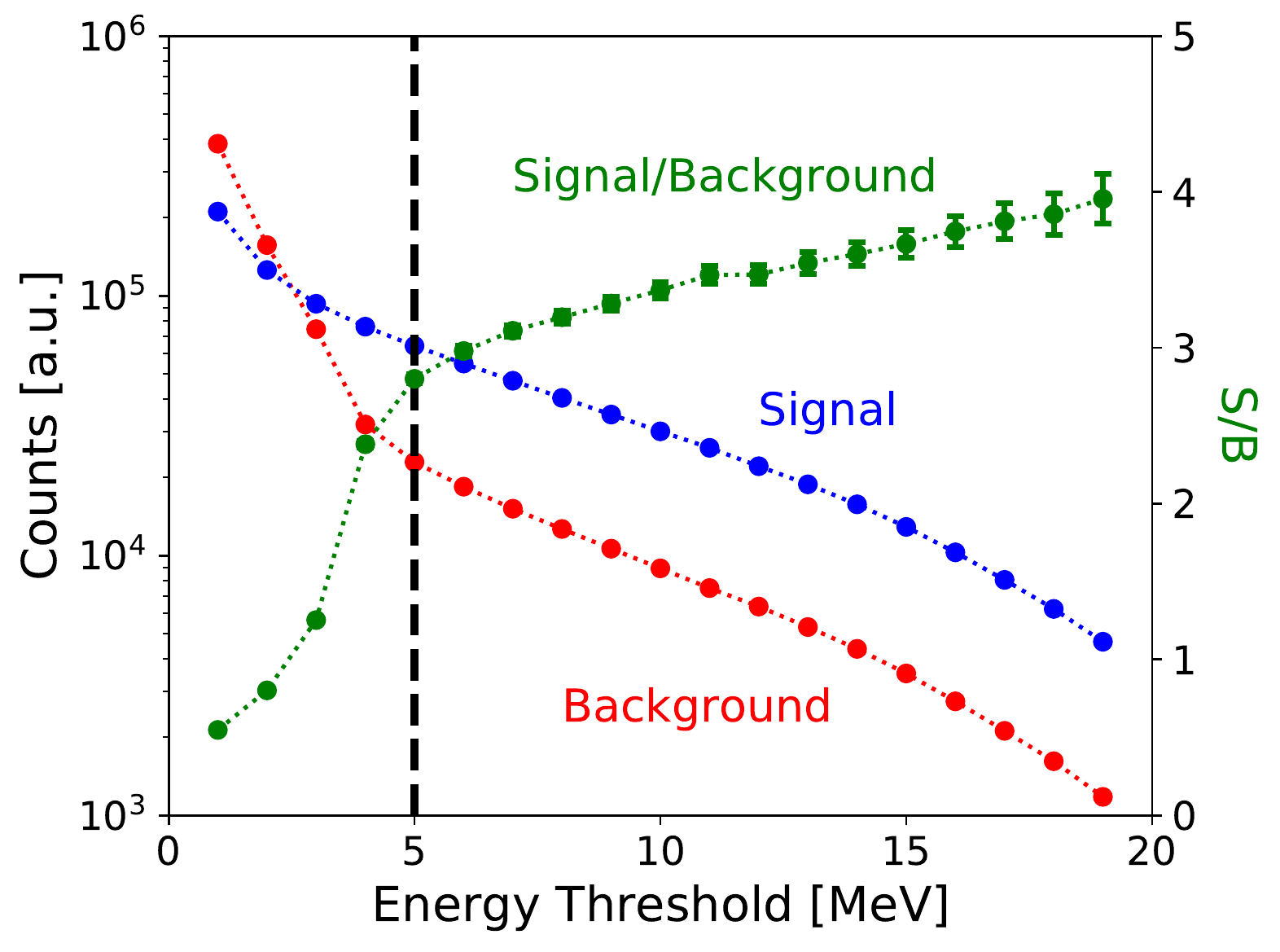}
	\caption{Signal counts (blue) and accidental background counts
          (red) in the neutron signal region of $200 \le p_n \le 600$ \si{\MeV/\clight}. The signal-to-background ratio (green) shown with a possible minimum energy deposition cut of 5 \si{\MeV}.}
	\label{fig:signalbackground}
\end{figure}

The minimum energy deposition for neutron signals is optimized using the neutron
signal-to-background ratio, see Fig.~\ref{fig:signalbackground}. The
signal region includes times for neutrons with $200\le p_n \le600$
\si{\MeV/\clight}. The accidental background was estimated by
fitting a constant to the left of the photon peak (see
Fig.~\ref{fig:tof}). The signal is defined as the integral of the
spectrum over the signal region minus the scaled background.  
As the minimum energy deposition cut increases, the signal decreases,
but the signal to background ratio increases.  
With a 5-\si{\MeV} energy deposition,
the signal-to-background ratio is almost three. The optimal cut will
be somewhere between two and seven \si{\MeV} and will be determined by
minimizing the experimental uncertainties.

\subsubsection{Neutron efficiency}

\begin{figure}[tbh!]
	\centering
		\includegraphics[width=0.48\textwidth]{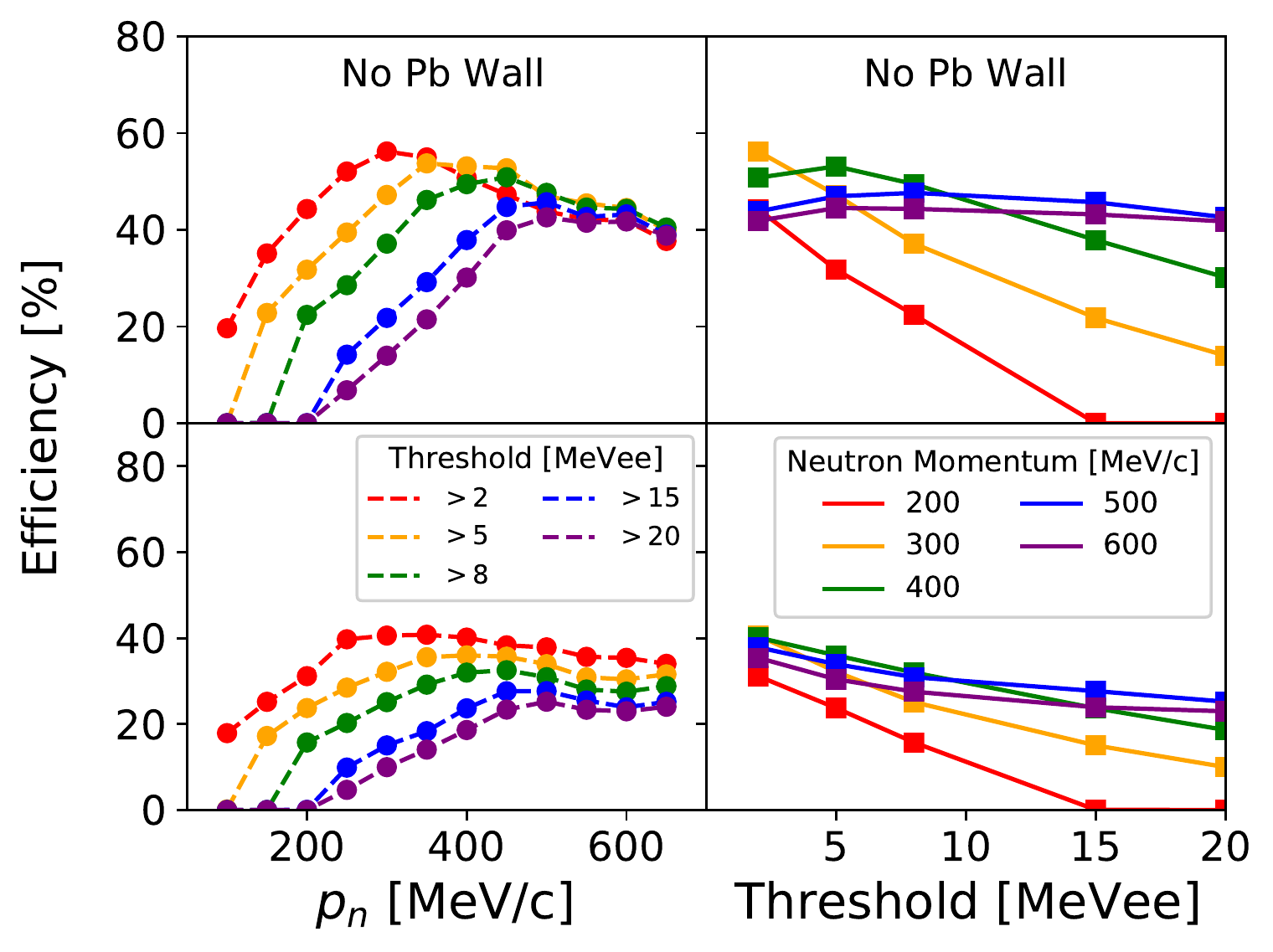}
	\caption{The efficiency of BAND for neutron detection as determined using a Geant4 simulation. (Left) Efficiency as a function of neutron momentum for various energy thresholds, without the effect of the lead wall upstream of BAND (top) and including the lead wall (bottom). (Right) Efficiency as
	a function of energy threshold for various neutron momenta without the effect of the lead wall upstream of BAND (top) and including the lead wall (bottom). }
	\label{fig:eff}
\end{figure}

The efficiency of BAND for neutron detection was studied using a Geant4 simulation, the results of which
are shown in Fig.~\ref{fig:eff}. As expected, the efficiency depends both on the energy threshold 
as well as the neutron momentum. 

The Geant4 study did not include the effects of the material between BAND
and the CLAS12 target, aside from the lead wall upstream of BAND. Neutrons of different fixed momenta were simulated according
to a flat solid-angle distribution, and the efficiency was calculated to be the number of events
registering only a single hit with energy deposition above threshold divided by the number of fired neutrons with 
trajectories hitting BAND. This efficiency is therefore an average over different incident angles
and hit positions in BAND.  This efficiency will be reduced by the effects of
neutron multiple
scattering in the material between BAND and the CLAS12 target. 

The simulation indicates that BAND can achieve an efficiency at or
above the design goal of above 35\% using thresholds between 2- and 
5~MeVee, even for higher-momentum neutrons. Running with higher thresholds to suppress background
may be possible without significantly compromising BAND's neutron
detection efficiency.


\section{Summary}
The Backward Angle Neutron Detector (BAND) of CLAS12 detects neutrons
  emitted at backward angles of $155$\si{\degree} to $175$\si{\degree}
  and momenta between $200$ and $600$ \si{\MeV/\clight}. It is
  positioned 3-\si{\meter} upstream of the target and consists of $18$
  rows and $5$ layers of $7.2$-\si{\centi\meter} by $7.2$-\si{\centi\meter} scintillator bars with PMT readout on both ends to
  measure time and energy deposition in the scintillator
  layers. Between the target and BAND there is a 2-\si{\centi\meter}
  thick lead wall followed by a 2-\si{\centi\meter} veto layer to
  suppress gammas and reject charged particles. 
  
  Timing calibrations were performed using a novel picosecond-pulsed laser system,
  along with cosmic and $ed \rightarrow e'n$ data. Energy calibrations utilized a wide range
  of radioactive sources.
  
  After timing and energy calibrations performed, the measured time
  resolution is below our design goal of 300 \si{\pico\s}, better than $250$~\si{\pico\s} (150~\si{\pico\s}) for energy depositions above 2~\si{\MeV} (5~\si{MeV}), yielding a momentum
  reconstruction resolution of $\delta p/p < 1.5$\%.

The expected neutron efficiency from simulation is $35$\%. In the
future, we will measure the efficiency using exclusive $ed \rightarrow e'pn$
events collected at Jefferson Lab.


\section*{Acknowledgements}
We thank Jefferson Lab's Hall B technical crew for the immense efforts during the BAND assembly, installation, and commissioning, especially Denny Insley, Calvin Mealer and Bob Miller. Thanks to Sergey Boyarinov, Ben Raydo, Chris Cuevas and Armen Stepanyan for providing crucial support in setting up the readout system.

We thank Bill Briscoe from George Washington University for generously providing PMTs and voltage dividers for the veto bars. We would like to thank MIT Bates for building the BAND support frame. 

We also would like to thank Hall B staff for their unwavering support, especially Stepan Stepanyan for setting up an assembly and testing area at Jefferson Lab and Eugene Pasyuk for organizing components. The development of the calibration, reconstruction, and analysis software benefited greatly from discussions, advice, and assistance by Daniel Carman, Raffaella De Vita, and Cole Smith. Finally, thanks to Run Group B of Hall B in Jefferson Lab, and our run-group coordinator, Silvia Niccolai, for their support during data-taking.

This work was supported by the U.S. Department of Energy, Office of Science, Office of Nuclear Physics under Award Numbers DE-FG02-94ER40818, DE-SC0020240, DE-FG02-96ER-40960, DE-FG02-93ER40771, and DE- AC05-06OR23177 under which Jefferson Science Associates operates the Thomas Jefferson National Accelerator Facility, the Israel Science Foundation under Grant Nos.\ 136/12 and 1334/16, the Pazy foundation, the Clore Foundation, and the Chilean FONDECYT grant No.\ 1201964.

\bibliography{band_nim_bib}

\begin{thebibliography}{10}
\expandafter\ifx\csname url\endcsname\relax
  \def\url#1{\texttt{#1}}\fi
\expandafter\ifx\csname urlprefix\endcsname\relax\def\urlprefix{URL }\fi
\expandafter\ifx\csname href\endcsname\relax
  \def\href#1#2{#2} \def\path#1{#1}\fi

\bibitem{Burkert:2020akg}
V.~Burkert, et~al., {The CLAS12 Spectrometer at Jefferson Laboratory}, Nucl.\
  Instrum.\ Meth.\ A 959 (2020) 163419.
\newblock \href {http://dx.doi.org/10.1016/j.nima.2020.163419}
  {\path{doi:10.1016/j.nima.2020.163419}}.

\bibitem{band-proposal}
O.~Hen, et~al.,
  \href{https://misportal.jlab.org/pacProposals/proposals/1159/attachments/93263/E12-11-003A.pdf}{In
  medium proton structure functions, src, and the emc effect}, JLab PAC
  Proposal (2015).
\newline\urlprefix\url{https://misportal.jlab.org/pacProposals/proposals/1159/attachments/93263/E12-11-003A.pdf}

\bibitem{Chatagnon:2020lwt}
P.~Chatagnon, et~al., {The CLAS12 Central Neutron Detector}, Nucl.\ Instrum.\
  Meth.\ A 959 (2020) 163441.
\newblock \href {http://dx.doi.org/10.1016/j.nima.2020.163441}
  {\path{doi:10.1016/j.nima.2020.163441}}.

\bibitem{band-laser}
A.~Denniston, et~al., {Laser Calibration System for Time of Flight Scintillator
  Arrays }, Nucl.\ Instrum.\ Meth.\ A 973 (2020) 164177.
\newblock \href {http://dx.doi.org/10.1016/j.nima.2020.164177}
  {\path{doi:10.1016/j.nima.2020.164177}}.

\bibitem{pmtR7724}
\text{Hamamatsu R7724}:
  \url{https://www.hamamatsu.com/resources/pdf/etd/R7723_R7724_R7725_TPMH1315E.pdf}.

\bibitem{pmt9214}
ET 9214,
  \url{http://et-enterprises.com/products/photomultipliers/product/p9214b-series}.

\bibitem{pmt9954}
ET 9954KB,
  \url{http://et-enterprises.com/products/photomultipliers/product/p9954b-series}.

\bibitem{scint-mat-ref}
\text{BC408 Scintillant}:
  \url{https://www.crystals.saint-gobain.com/sites/imdf.crystals.com//files/documents/bc400-404-408-412-416-data-sheet.pdf}.

\bibitem{3MESR}
3M Enhance Specular Reflector:
  \url{https://www.3m.com/3M/en_US/company-us/all-3m-products/~/3M-Enhanced-Specular-Reflector-3M-ESR-/?N=5002385+3293061534&rt=rud}.

\bibitem{hamapmts}
Overview over Hamamatsu PMTs:
  \url{https://www.hamamatsu.com/resources/pdf/etd/High_energy_PMT_TPMZ0003E.pdf}.

\bibitem{uvglue}
\text{DYMAX Ultra Light-Weld 3-20262},
  \url{http://www.dymax.com/images/pdf/pds/3-20262.pdf}.

\bibitem{softglue}
\text{MOMENTIVE RTV615 silicone rubber compound},
  \url{https://www.momentive.com/en-us//products/tds/rtv615?productId=64616a84-08ff-4372-a2a2-a58eb45be627}.

\bibitem{Fair:2020yfx}
R.~Fair, et~al., {The CLAS12 superconducting magnets}, Nucl.\ Instrum.\ Meth.\
  A 962 (2020) 163578.
\newblock \href {http://dx.doi.org/10.1016/j.nima.2020.163578}
  {\path{doi:10.1016/j.nima.2020.163578}}.

\bibitem{thorlabsfoil}
\text{Thorlabs black Aluminium foil},
  \url{https://www.thorlabs.com/thorproduct.cfm?partnumber=BKF12}.

\bibitem{teem_laser}
\text{Teem Photonics}:
  \url{https://www.teemphotonics.com/laser/microchip-laser-355-nm-1e-4khz-600ps}.

\bibitem{attenuator}
\text{OZ Optics}: \url{https://www.ozoptics.com/products/attenuators.html}.

\bibitem{caen-hvframe}
CAEN SY4527: \url{https://www.caen.it/products/sy4527/}.

\bibitem{caen-hvcard}
CAEN A1535SN:
  \url{http://www.caen-group.com/servlet/checkCaenManualFile?Id=9315}.

\bibitem{fadc-manual}
\href{https://www.jlab.org/Hall-B/ftof/manuals/FADC250UsersManual.pdf}{JLab 250
  MHz FADC manual}.
\newline\urlprefix\url{https://www.jlab.org/Hall-B/ftof/manuals/FADC250UsersManual.pdf}

\bibitem{caen-tdc}
CAEN VX1190A: \url{https://www.caen.it/products/v1190a-2esst/}.

\bibitem{clas12-daq}
S.~Boyarinov, et~al., {The CLAS12 Data Acquisition System}, Nucl.\ Instrum.\
  Meth.\ A 966 (2020) 163698.
\newblock \href {http://dx.doi.org/10.1016/j.nima.2020.163698}
  {\path{doi:10.1016/j.nima.2020.163698}}.

\bibitem{clas12-trigger}
B.~Raydo, et~al., {The CLAS12 Trigger System}, Nucl.\ Instrum.\ Meth.\ A 960
  (2020) 163529.
\newblock \href {http://dx.doi.org/10.1016/j.nima.2020.163529}
  {\path{doi:10.1016/j.nima.2020.163529}}.

\bibitem{caen-logicboard}
CAEN V1495: \url{https://www.caen.it/products/v1495/}.

\bibitem{tedlarfoil}
\text{Dupont Tedlar foil},
  \url{https://www.dupont.com/content/dam/dupont/products-and-services/membranes-and-films/pvf-films/documents/DEC_Tedlar_GeneralProperties.pdf}.

\bibitem{comptonedge}
M.~J. Safari, F.~Abbasi~Davani, H.~Afarideh, {Differentiation method for
  localization of Compton edge in organic scintillation detectors }\href
  {http://arxiv.org/abs/1610.09185} {\path{arXiv:1610.09185}}.

\end{thebibliography}

\end{document}